\pdfoutput=1
\documentclass[twocolumn,prd,preprintnumbers,aps,amssymb]{revtex4-1}
\usepackage{dcolumn}
\usepackage{graphicx}
\usepackage{amsmath}
\usepackage{bm}
\usepackage[colorlinks=true, pdfstartview=FitV, linkcolor=red, citecolor=blue, urlcolor=blue, pdftitle={Renormalized Linear Kinetic Theory as Derived from Quantum Field Theory --- a novel diagrammatic method for computing transport coefficients ---},pdfauthor={Yoshimasa Hidaka and Teiji Kunihiro},pdfsubject={}, pdfkeywords={transport coefficient, viscosity, quark gluon plasma}]{hyperref}

\newcommand{\arXiv}[1]{\href{http://arxiv.org/abs/#1}}
\newcommand{\mydoi}[1]{\href{http://dx.doi.org/#1}}
\newcommand{\IRCutoff}{\kappa}
\newcommand{\DR}{D^R}
\newcommand{\DA}{D^A}
\newcommand{\DFree}{D_{F} }
\newcommand{\DRFree}{D^{R}_{F}}
\newcommand{\DAFree}{D^{A}_{F}}
\newcommand{\bareThreePoint}{\varGamma^\text{(0)}}
\newcommand{\fourPointVertex}{\mathcal{T}}
\newcommand{\fourPointVertexA}{\fourPointVertex^{(A)}}
\newcommand{\fourPointVertexB}{\fourPointVertex^{(B)}}
\newcommand{\re}{\mathrm{Re}\,}
\newcommand{\im}{\mathrm{Im}\,}

\newcommand{\transportCoefficient}{\sigma}

\def\anp#1#2#3{Annals Phys. {\bf #1}, #2 (#3)}
\def\arnps#1#2#3{Ann.\ Rev.\ Nucl.\ Part.\ Sci.\  {\bf #1}, #2 (#3)}

\def\ibid#1#2#3{{\it ibid.} {\bf #1}, #2 (#3)}
\def\ijm#1#2#3{Int. Jour. Mod. Phys. {\bf #1}, #2 (#3)}
\def\jhep#1#2#3{Jour. High Energy Phys. {\bf #1}, #2 (#3)}

\def\npa#1#2#3{Nucl. Phys. A {\bf #1}, #2 (#3)}
\def\npb#1#2#3{Nucl. Phys. B {\bf #1}, #2 (#3)}

\def\plb#1#2#3{Phys. Lett. B {\bf #1}, #2 (#3)}
\def\pr#1#2#3{Phys. Rev. {\bf #1}, #2 (#3)}
\def\prc#1#2#3{Phys. Rev. C {\bf #1}, #2 (#3)}
\def\prd#1#2#3{Phys. Rev. D {\bf #1}, #2 (#3)}
\def\prl#1#2#3{Phys. Rev. Lett. {\bf #1}, #2 (#3)}
\def\phr#1#2#3{Phys. Rep. {\bf #1}, #2 (#3)}

\def\ptps#1#2#3{Prog. Theor. Phys. Suppl. {\bf #1}, #2 (#3)}

\begin{document}
\preprint{KUNS-2301}
\title{Renormalized Linear Kinetic Theory as Derived from Quantum Field Theory \\
--- a novel diagrammatic method for computing transport coefficients ---}
\author{Yoshimasa Hidaka}
\author{Teiji Kunihiro}
\affiliation{
Department of Physics, Kyoto University, Sakyo-ku, Kyoto 606-8502, Japan}
\begin{abstract}
We propose a novel diagrammatic method for computing transport coefficients 
in relativistic quantum field theory. 
The self-consistent equation for summing the diagrams 
with pinch singularities has a form of  a linearized kinetic equation as usual,
but our formalism enables us to incorporate higher-order corrections of the coupling 
 systematically for the first time. 
Furthermore, it is clarified that the higher-order corrections are nicely summarized
into that of the vertex function, spectral function, and collision term. 
We identify the diagrams up to the next-to-next-leading order corrections
in the weak coupling expansion of $\phi^4$ theory,
which is a difficult task in kinetic approaches and other diagrammatic methods.
\end{abstract}

\maketitle
\section{Introduction}
The experimental results of heavy-ion collisions at relativistic heavy ion collider
 (RHIC) brought 
remarkable results \cite{whitepaper,strong,strong_rdp,hydro}. 
Among others, elliptic flow in peripheral collisions suggests that the ratio 
of the shear viscosity to 
the entropy, $\eta$/s, is smaller than the ever observed in other 
systems~\cite{hydro,Csernai:2006zz}.
Transport coefficients such as the shear $\eta$ and the bulk $\zeta$ viscosities 
reflect dynamical 
properties slightly away from the thermal equilibrium, while pressure, 
entropy, susceptibility, etc., 
do static properties.
In the kinetic theory, the shear viscosity is proportional to 
the mean free path of quasiparticles of a long life;
a short mean free path corresponds to strongly interacting 
matter~\cite{PhysicalKinetics}, 
which is the reason why the quark gluon plasma (QGP) at RHIC is called 
a strongly-coupled quark gluon plasma (sQGP)~\cite{strong}. 
In any event, the RHIC experiments and the subsequent analyses
prompted a great interest in the relativistic quantum field
theory for the transport coefficients or non-equilibrium dynamics in general.

Formally, the transport coefficients are expressed 
by the so called Kubo formula~\cite{KuboFormula} 
in the framework of linear response theory even for the relativistic system;
 e.g., 
the shear and the bulk viscosities are given by
\begin{align}
\eta &= \frac{1}{10}\lim_{\omega\to0}\frac{1}{\omega}\,\im
\int d^4x \,e^{i\omega t} i\theta(t)\langle[\pi_{ij}(x),\pi^{ij}(0)]\rangle \,,  \label{eq:KuboFormulaShear} \\
\zeta &= \lim_{\omega\to0}\frac{1}{\omega}\,\im
\int d^4x\, e^{i\omega t} \,
i \theta(t)\langle [\mathcal{P}(x),\mathcal{P}(0)]\rangle \, ,
\label{eq:KuboFormulaBulk}
\end{align}
respectively, where $\mathcal{P}(x)=-{T^{i}}_{i}(x)/3$ and 
$\pi_{ij}(x)=T_{ij}(x)+g_{ij}\mathcal{P}(x)$. 
$T_{\mu\nu}(x)$ is the energy-momentum tensor.
We work in Minkowski space with a metric,  $g_{\mu\nu}=\text{diag}\,(1,-1,-1,-1)$.
Although concisely expressed by the Kubo formula,
 the computation of  the transport coefficients 
is not a simple task in practice.

Lattice QCD simulation in principle provides us with a powerful nonperturbative method,
and it is natural that there are attempts to apply it 
to calculate transport coefficients at vanishing chemical 
potential~\cite{lattice1,Huebner:2008as,Meyer:2009jp}.
However, 
one must notice that the calculation of the energy-momentum tensor on the lattice is
not by far straightforward because a more careful analysis would be necessary 
of the low frequency structure
of the spectral function and the relevant relaxation time scales which 
are to be extracted from the imaginary-time correlation function,
as warned in~\cite{Huebner:2008as}.  

Recently,  calculations relying on gauge/gravity correspondence have become
popular, in which the retarded correlation function can be calculated 
as the absorbed cross section  of the black hole~\cite{susy1,susy2}.
One of the most important results from such an approach
is that the ratio 
of the shear viscosity to the entropy density in  ${\cal N}=4$-supersymmetric 
gauge theories
is as small as $1/4 \pi$
with infinite number of colors and strong coupling, and this value
 is conjectured 
to be a universal lower bound \cite{susy1,susy2}. 
 
Even with the recent development of the fancy methods for calculating 
the transport coefficients,
the most reliable method with  the sound basis in the relativistic quantum
field theory would be
diagrammatic ones based on the loop expansion.
One of the most important and difficult part of such a method is, however, 
how to implement a systematic way to
deal with the so called ``pinch singularity'',
which would make naive perturbation theory based on the loop expansion breaks down.
In fact, this difficulty could  be overcome by a resummation of  specific ladder diagrams.
Although restricted to the leading order 
of a coupling constant or large-$N$ expansion,
several methods for such a resummation have been 
proposed~\cite{Jeon,diagram,2PI,3PI,Gagnon}, and
 their outcome has been shown to be equivalent to the results 
as obtained using the relativistic kinetic 
or Boltzmann equation within
 the leading order of a coupling constant~\cite{Jeon,Gagnon}.
We here remark that 
direct applications of  relativistic kinetic theories are
also done to compute the transport coefficients
in the hadronic~\cite{hadronic},  
quark-gluon plasma~\cite{transport,amy}  
and the `semi-QGP' phases~\cite{semiQGP}.

The merit of the diagrammatic methods based on the loop expansion 
over the direct use of 
the Boltzmann equation
is the potential ability to take into account the higher-order effects 
of the coupling constant systematically,  in principle.
As far as we know, 
there were, however, only few works which were able to examine
 the next leading order for a relativistic system,
 still being confined to a scalar theory~\cite{nextLeading1,nextLeading2}.
The major technical difficulty for going beyond the leading order 
of the coupling constant 
is to classify diagrams order by order 
in the coupling constant, 
because the loop expansion no longer corresponds to the coupling expansion 
at finite temperature and/or density.

The purpose of this paper is to give a formulation for a novel
resummation method 
in which  transport coefficients  can be calculated systematically  
for any order of the coupling constant 
 beyond the leading order for a relativistic system.
Our method is motivated by the Eliashberg theory 
 for a nonrelativistic system~\cite{Eliashberg:1962} like a Fermi-liquid, 
although it relies on a cumbersome analytic continuation
inherent in the imaginary-time formalism adopted by  Eliashberg. 
We extend and reformulate the resummation method developed by Eliashberg 
to relativistic quantum field theory at high temperature. 
One of our ideas is to employ the real-time formalism to avoid the complicated 
analytical continuation,
 and thus make the calculational procedure transparent.
 The real-time formalism is also found convenient 
to identify diagrams corresponding to the pinch singularity, 
 as will be discussed in Sec.~\ref{sec:transportCoefficient}.

The paper is organized as follows:
In Sec.~\ref{sec:quasiParticles}, we briefly review why the difficulty arises 
when computing the transport coefficients.
In Sec.~\ref{sec:realTimeFormalism}, we review the real-time formalism in 
the $R/A$ basis 
that is useful to identify the dominant diagrams in the Green function 
corresponding to the transport coefficient.
In Sec.~\ref{sec:transportCoefficient}, the computation method for 
transport coefficients are discussed, 
and we show that the resummation equation has a form similar 
to a linearized Boltzmann equation. 
In Sec.~\ref{sec:Conclusion}, we will conclude the paper with some outlook.

\section{Pinch singularity and quasiparticles} \label{sec:quasiParticles}

The difficulty in computing transport coefficients  in Eqs.~(\ref{eq:KuboFormulaShear}) 
and (\ref{eq:KuboFormulaBulk})
arises from the infrared limit, $\omega\to0$, of the Green function corresponding 
to  a long time scale.
In such long time scales, a lot of microscopic scatterings occur.
In a diagrammatic method, such scatterings are expressed by multiple loop diagrams. 
The question is what kind of diagrams are dominating and should be resummed.
A particle excitation of a long life is called a quasiparticle, and
as long as the quasiparticle description is valid for the system under consideration,
thermal excitations of the system will be 
well described solely  by the quasiparticle excitations.
Such an excitation is expressed by a product of the retarded and advanced Green functions 
of the quasiparticles,
which product is found to become divergent in the infinite lifetime limit.
This divergence or singularity is nothing but
the pinch singularity that we mentioned in Introduction.
 
To see this more concretely, 
 let us take the $\phi^4$ theory, in which the one-particle retarded and
advanced propagator  read
\begin{equation}
\begin{split}
\DR(k)&=\frac{-1}{k^{2}-m^{2}-\re\varPi_{R}(k)-i \,\im\varPi_{R}(k)} \\
&={\DA}^{*}(k)\,,
\end{split}
\label{eq:dressedPropagaotor}
\end{equation}
where $m$ is the mass of the scalar particle, and $\varPi_{R}(k)$ is the retarded self-energy.
These propagator can be approximated by a sum of poles corresponding to quasiparticle excitations
and thus  we have
\begin{align}
\DR(k)&\simeq \sum_{n}\frac{-z_{n}(\bm{k})}{k^{0}- \epsilon_{n}(\bm{k})+i\gamma_{n}(\bm{k})} \,, \\
\DA(k)&\simeq \sum_{n}\frac{-z^*_{n}(\bm{k})}{k^{0}- \epsilon_{n}(\bm{k})-i\gamma_{n}(\bm{k})} \,,
\end{align}
where $\epsilon_n(\bm{k})$ is the energy of the quasiparticle, $z_{n}(\bm{k})$ 
the renormalization function, and $\gamma_{n}(\bm{k})>0$ the damping rate.
At weak coupling and high temperature ($m\ll\lambda T$),
$\epsilon_n(\bm{k})\simeq\pm\sqrt{\bm{k}^2+m_T^2}$, 
with the thermal mass defined in Eq.~(\ref{eq:thermalMass}) bellow,
 and $z_n(\bm{k})\simeq1/(2|\epsilon_n(\bm{k})|)$. 
 The damping rate owing to $2\to2$ scattering  is of order $\lambda^2 T$.
In the naive perturbation theory,
one would encounter a product of the retarded and the advanced propagators, $\DR(p+k)\DA(k)$, 
which has the following anomalous behavior at small $p$ under the quasiparticle approximation:
\begin{equation}
\begin{split}
&\DR(p+k)\DA(k)\\
&\quad\simeq 2\pi i\sum_{n} \frac{|z_{n}
(\bm{k})|^{2}}{p^{0}-\bm{v}_{n}\cdot\bm{p}+2i\gamma_{n}(\bm{k})}
\delta\bigl(k^{0}-\epsilon_{n}(\bm{k})\bigr) \,,
\end{split}
\label{eq:pinchsing}
\end{equation}
where $\bm{v}_{n}=d\epsilon_{n}(\bm{k})/d\bm{k}$ is the velocity of the quasiparticle.
Equation~(\ref{eq:pinchsing}) shows that when the damping rate vanishes, i.e., $\gamma_n(\bm{k})=0$, 
the product diverges at $p=0$;
 this {\em is} the pinch singularity as mentioned above. 
And one sees that the product of the retarded and the advanced propagator
 with the common momentum must be summed over to have a sensible result.

The need for resummation can be seen even apart from the quasiparticle approximation.
The product of the propagators at $p=0$ exactly becomes
\begin{equation}
\lim_{p\to0}\DR(p+k)\DA(k) = \frac{\rho(k)}{-2\,\im\varPi_R(k)} \,,
\label{eq:pairPropagatorAtP=0}
\end{equation}
where $\rho(k)$ is the spectral function defined by
\begin{equation}
\begin{split}
\rho(k) &\equiv  2 \,\im \DR(k)\\
& = \frac{-2\,\im \varPi_{R}(k)}{\bigl(k^{2}-m^{2}-\re\varPi_{R}(k)\bigr)^{2}+\bigl(\im \varPi_{R}(k)\bigr)^{2}} \,.
\end{split}
\end{equation}
One sees that Eq.~(\ref{eq:pairPropagatorAtP=0}) diverges
 when the imaginary part of the self-energy vanishes.
Therefore, 
we have to employ the dressed propagator, Eq.~(\ref{eq:dressedPropagaotor}), to avoid the singularity.

\section{real-time formalism}
\label{sec:realTimeFormalism}
\begin{figure}
\begin{center}
\includegraphics[width=0.4\textwidth]{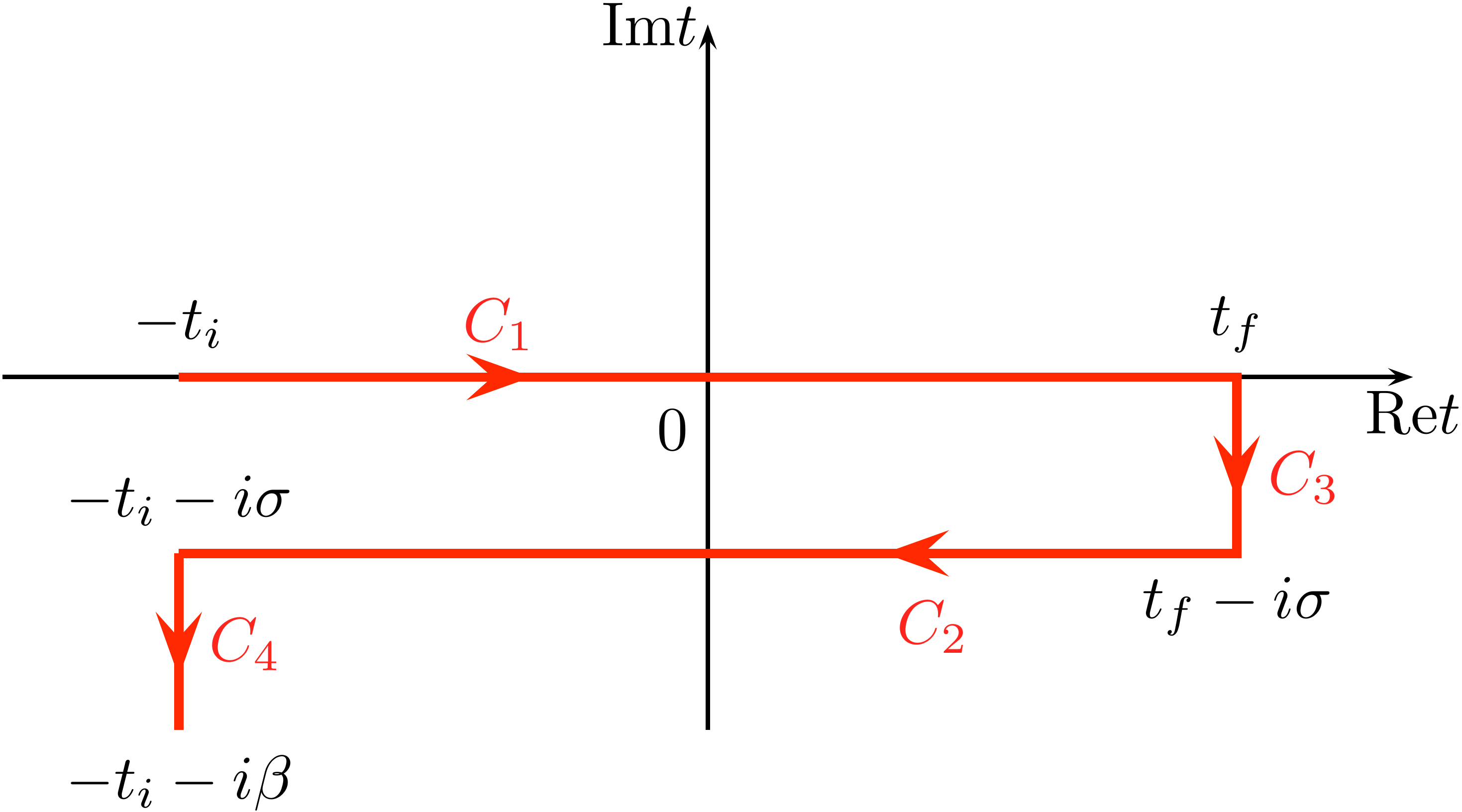}
\end{center}
\caption{A complex-time path in the real-time formalism.}
\label{fig:timepath}
\end{figure}
In this section, we briefly review the real-time formalism~\cite{lebellac}. 
For simplicity, we consider the $\phi^{4}$ theory. 
The Lagrangian has the form,
\begin{equation}
\mathcal{L}=\frac{1}{2}(\partial_{\mu}\phi)^{2}-\frac{m^{2}}{2}\phi^{2}- \frac{\lambda}{4!}\phi^{4} \,.
\end{equation}
The real-time formalism is formulated on a complex-time path shown in Fig.~\ref{fig:timepath}.
The fields on $C_1$ and $C_2$ are called type-1 field, $\phi^{1}$, and type-2 field, $\phi^2$, respectively.
The partition function is
\begin{equation}
\begin{split}
\mathcal{Z}_{12}=&\int \mathcal{D}\phi^1\mathcal{D}\phi^2\exp\Bigl[-\frac{1}{2}\int d^{4}x d^{4}x'\\
&\quad\times
\phi^a(x)D^{-1}_{F;ab}(x,x')\phi^b(x')\\
&\qquad- i \int d^{4}x \bigl(\mathcal{V}(\phi^{1}) -\mathcal{V}(\phi^{2}) \bigr)  \Bigr] \,.
\end{split}
\end{equation}
Although there are contributions from the paths $C_3$ and $C_4$ in general, 
they are factorized in the limit of $t_i, t_f\to \infty$, 
and are irrelevant as long as correlation functions are concerned~\cite{lebellac}.
The free propagators in momentum space read
\begin{equation}
\begin{split}
\DFree^{11} &=  \bigl(1+n(k^{0})\bigr)D_{F}(k)+n(k^{0})D_{F}^*(k) \,,\\
\DFree^{12} &= 
\,e^{\sigma k^{0}}\bigl(n(k^{0})+\theta(-k^{0})\bigr)\bigl(D_{F}(k)+D_{F}^*(k)\bigr) \,,\\
\DFree^{21}&=e^{-\sigma k^{0}}\bigl(n(k^{0})+\theta(k^{0})\bigr)\bigl(D_{F}(k)+D_{F}^*(k)\bigr) \,,\\
\DFree^{22}&= \bigl(1+n(k^{0})\bigr)D_{F}^*(k)+n(k^{0})D_{F}(k) \,,
\end{split}
\label{eq:12basis}
\end{equation}
where the $\sigma$ is the parameter characterizing the time path in Fig.~\ref{fig:timepath},
$\DFree(k)$ is  the free Feynman propagator defined by
\begin{equation}
\DFree(k)=\frac{i}{k^{2}-m^{2} + i\epsilon} \,,
\end{equation}
and $n(k^{0})$ is the Bose-Einstein distribution function,
\begin{equation}
n(k^{0})=\frac{1}{e^{\beta |k^{0}|}-1} \,,
\end{equation}
with inverse temperature $\beta=1/T$.
We note that any physical observable is independent of the choice of  the path, i.e.,
$\sigma$.

For later convenience, let us rewrite Eq.~(\ref{eq:12basis}) in terms of the retarded and 
advanced propagators,
\begin{equation}
\begin{split}
\DFree^{11}(k) & =  \bigl(1+f(k^{0})\bigr)\bigl(-i \DRFree(k)\bigr)+f(k^{0})\bigl(i\DAFree(k)\bigr) \,,\\
\DFree^{12}(k)& = 
e^{\sigma k^{0}} f(k^{0})\bigl(-i\DRFree(k)+i\DAFree(k)\bigr) \,,\\
\DFree^{21}(k) & = 
e^{-\sigma k^{0}}\bigl(1+f(k^{0})\bigr)\bigl(-i\DRFree(k)+i\DAFree(k)\bigr) \,,\\
\DFree^{22}(k) & = f(k^{0})\bigl(-i \DRFree(k)\bigr)+\bigl(1+f(k^{0})\bigr)\bigl(i\DAFree(k)\bigr) \,,
\end{split}
\label{eq:RAbasis}
\end{equation}
where the free retarded and advanced propagators are defined by
\begin{align}
\DRFree(k)&\equiv\frac{-1}{k^{2}-m^{2}+i\epsilon k^{0}} \,, \\
\DAFree(k)&\equiv\frac{-1}{k^{2}-m^{2}-i\epsilon k^{0}} \,,
\end{align}
and, $f(k^{0})$ is defined as
\begin{equation}
f(k^{0})\equiv\frac{1}{e^{\beta k^{0}}-1} \,.
\end{equation}
These propagators satisfy $\DRFree(-k)={\DRFree}^*(k)=\DAFree(k)$, 
which relation also holds for the full propagators (see Appendix~\ref{sec:identities}.)
The relation between the Bose-Einstein distribution function and $f(k^{0})$ is $n(k^{0})=f(|k^{0}|)$.

\subsection{$R/A$ basis}

In this subsection, we introduce a useful basis called ``$R/A$ basis'' for 
computing transport coefficients~\cite{Aurenche:1991hi,vanEijck:1994rw}.
In the original representation introduced by Aurenche and Becherrawy~\cite{Aurenche:1991hi}, 
the propagators are diagonal, and the diagonal components are the retarded and advanced propagators.
In the present work, we rather employ the representation in which the diagonal components 
are zero while the off-diagonal components are not \cite{vanEijck:1994rw},
 instead of the original $R/A$ basis (we also refer this as $R/A$ basis in this paper.)
We find that this basis is useful  not only for a comparison
 of the  real-time correlation function obtained in the real-time formalism with that 
in the imaginary-time but also for identification of pinch singularities.

Let us define a rotation matrix for converting the basis Eq.~(\ref{eq:12basis})
 (we refer this basis as ``standard basis'') to the $R/A$ basis in momentum space as
\begin{equation}
\phi^\alpha(k)= {U^\alpha}_a(k) \phi^a(k)  \,,
\end{equation}
where the Greek index denotes $R$ or $A$, and the Latin index denotes $1$ or $2$. 
We use Einstein notation, i.e.,
if an index appears twice in a single term, once as a superscript and once as subscript, 
a summation is assumed over all of its possible values.
We also define the metric for the standard basis as $g_{ab}=g^{ab}\equiv\mathrm{diag}(1,-1)$; 
then, the metric in the $R/A$ basis is given by $g_{\alpha\beta} = g_{ab}{U^{a}}_{\alpha}(k){U^{b}}_{\beta}(-k)$, 
of which explicit form is given below.
The inverse matrix ${U^b}_\alpha(k)$ is defined by $\delta^b_a={U^b}_\alpha(k){U^\alpha}_a(k)$.
An $n$-point function transforms as a tensor:
\begin{align}
&\varGamma_{\alpha\beta\cdots}(k_1,k_2,\cdots) \notag\\
&\qquad=
\varGamma_{ab\cdots}(k_1,k_2,\cdots)
{U^a}_\alpha(k_1) {U^b}_\beta(k_2)\cdots  \,,\\
&D^{\alpha\beta\cdots}(k_1,k_2,\cdots)\notag\\
&\qquad=D^{ab\cdots}(k_1,k_2,\cdots){U^\alpha}_a(k_1)  {U^\beta}_b(k_2)\cdots \,.
\end{align}
In particular, the propagator or  two-point function transforms as
\begin{equation}
D^{ab}(k)={U^a}_\alpha(k) {U^b}_\beta(-k) D^{\alpha\beta}(k) \,.
\end{equation}
In the $R/A$ basis, $D^{\alpha\beta}(k)$ is chosen so that the diagonal part vanishes:
\begin{equation}
\begin{split}
D^{\alpha\beta}(k)&= 
\begin{pmatrix}
D^{RR}(k) & D^{RA}(k) \\
D^{AR}(k)  & D^{AA}(k)
\end{pmatrix}\\
&=
\begin{pmatrix}
0 & -i \DR(k) \\
-i \DA(k) & 0
\end{pmatrix} \,.
\end{split}
\label{eq:R/AbasisCondition}
\end{equation}
The most general  form of ${U^a}_\alpha(k)$ satisfying Eq.~(\ref{eq:R/AbasisCondition})
is obtained through comparison with Eq.~(\ref{eq:RAbasis}):
\begin{equation}
\begin{split}
{U^a}_\alpha(k)&=
\begin{pmatrix}
{U^1}_{R}(k) & {U^{1}}_{A}(k) \\
{U^{2}}_{R}(k)  & {U^{2}}_{A}(k)
\end{pmatrix}\\
&=\begin{pmatrix}
f(k^{0})e^{\sigma k^{0}}c(k) & e^{(\sigma-\beta) k^0}/c(-k) \\
f(k^{0})c(k) & 1/c(-k)
\end{pmatrix} \,,
\end{split}
\end{equation}
where $c(k)$ is an arbitrary function. Here, we simply fix $c(k)$ to be constant, $c(k)=1$,
while the $\sigma$ is kept to be a free parameter.
Then, 
\begin{equation}
{U^a}_\alpha(k)
=\begin{pmatrix}
f(k^{0})e^{\sigma k^{0}} & e^{(\sigma-\beta) k^0} \\
f(k^{0}) & 1
\end{pmatrix} \,,
\end{equation}
and the inverse is 
\begin{equation}
\begin{split}
{U^\alpha}_a(k)=&
\begin{pmatrix}
{U^R}_{1}(k) & {U^R}_{2}(k) \\
{U^A}_{1}(k)  & {U^A}_{2}(k)
\end{pmatrix}\\
=&
\begin{pmatrix}
e^{(\beta-\sigma)k^{0}}   & -1 \\
-\bigl(1+f(k^{0})\bigr)e^{-\sigma k^{0}}  & \bigl(1+ f(k^{0})\bigr) 
\end{pmatrix}\,.
\end{split}
\end{equation}
The metric in the $R/A$ basis is 
\begin{equation}
g_{\alpha\beta}=g_{ab}{U^a}_\alpha(k){U^b}_\beta(-k)
=
\begin{pmatrix}
0 &  1 \\
1 & 0
\end{pmatrix}
=g^{\alpha\beta} \,,
\end{equation}
which is independent of $\sigma$.
The propagator can be rewritten by using the metric $g^{\alpha\beta}$: 
\begin{equation}
D^{\alpha\beta}(k)=-ig^{\alpha\beta}D^{\alpha}(k) \,,
\label{eq:propagator}
\end{equation}
 with no sum over $\alpha$. Here we define the Feynman rule for the propagators with arrows:
 \begin{align}
\quad\parbox{2.3cm}{\vspace{0.3cm}\includegraphics[width=0.125\textwidth]{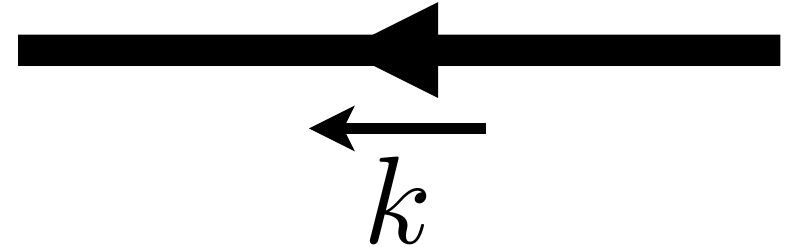}}
&= -i \DR(k)\,, \label{eq:feynmanRuleDR}\\
\quad\parbox{2.3cm}{\vspace{0.3cm}\includegraphics[width=0.125\textwidth]{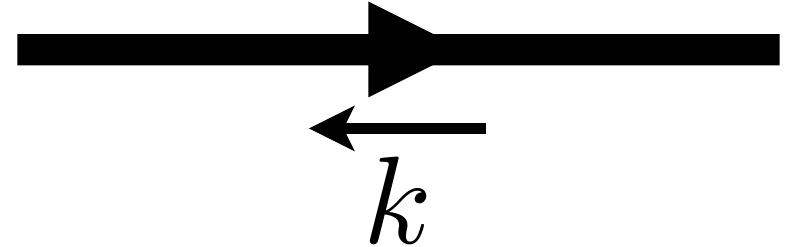}}
&= -i \DA(k) = -i \DR(-k)\,.
\label{eq:propagator2}
\end{align}
The propagator is identified with $D^R(k)$ or $D^A(k)$, if the arrow is parallel or anti-parallel to the momentum $k$, respectively.
The four-point vertex transforms  as
\begin{equation}
\begin{split}
&\lambda_{\alpha\beta\gamma\delta}(k_{1},k_{2},k_{3},k_{4})\\
&\quad= 
\lambda_{abcd}{U^{a}}_{ \alpha }(k_{1}){U^{b}}_{ \beta }(k_{2}){U^{c}}_{ \gamma }(k_{3})
{U^{d}}_{ \delta }(k_{4}) \,,
\end{split}
\end{equation}
where $\lambda_{abcd}=+\lambda$ if $a=b=c=d=1$, and $\lambda_{abcd}=-\lambda$ if 
 $a=b=c=d=2$, otherwise $\lambda_{abcd}=0$. The vertex satisfies the energy-momentum conservation, $k_{1}+k_{2}+k_{3}+k_{4} = 0$. After simple calculations, one finds
\begin{equation}
\begin{split}
&\lambda_{\alpha\beta\gamma\delta}(k_{1},k_{2},k_{3},k_{4})\\
&\quad= \lambda\Bigl[
\bigl(1+f(k^{0}_{1})\bigr)^{\delta_{\alpha R}}\bigl(1+f(k^{0}_{2})\bigr)^{\delta_{\beta R}}\\
&\qquad\qquad\times\bigl(1+f(k^{0}_{3})\bigr)^{\delta_{\gamma R}}\bigl(1+f(k^{0}_{4})\bigr)^{\delta_{\delta R}} \\
&\quad\qquad-\bigl(f(k^{0}_{1})\bigr)^{\delta_{\alpha R}}\bigl(f(k^{0}_{2})\bigr)^{\delta_{\beta R}}\bigl(f(k^{0}_{3})\bigr)^{\delta_{\gamma R}}
\bigl(f(k^{0}_{4})\bigr)^{\delta_{\delta R}} \Bigr]
\,.
\end{split}
\end{equation}
The first and second terms in RHS corresponds to the loss and the gain terms, respectively.
The four-point vertex generally contains sixteen combinations; however,  thanks to the symmetry of
the $\phi^{4}$ theory, they reduce to the following five vertices:
\begin{align}
\parbox{2.cm}{\includegraphics[width=0.1\textwidth]{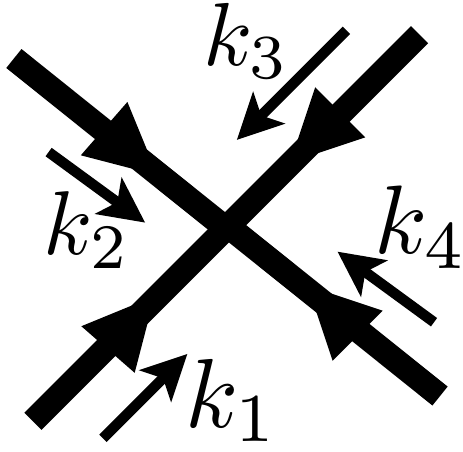}}&=
-i\lambda_{RRRR}(k_{1},k_{2},k_{3},k_{4})
=0 \,,\\
\parbox{2.cm}{\includegraphics[width=0.1\textwidth]{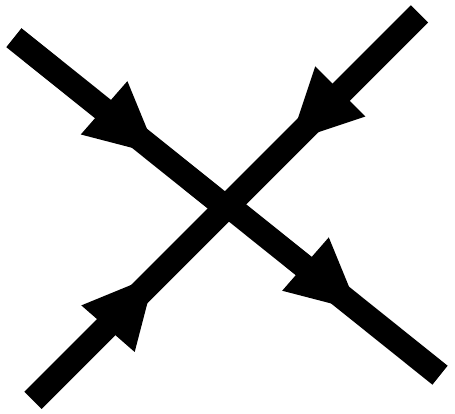}}&=-i\lambda_{RRRA}(k_{1},k_{2},k_{3},k_{4})\notag\\
&= -i\lambda\Bigl[\bigl(1+f(k^{0}_{1})\bigr)\bigl(1+f(k^{0}_{2})\bigr)\bigl(1+f(k^{0}_{3})\bigr) \notag\\&
\qquad\qquad\qquad\qquad- f(k^{0}_{1})f(k^{0}_{2})f(k^{0}_{3}) \Bigr] \,,
\end{align}
\begin{align}  
\parbox{2.cm}{\includegraphics[width=0.1\textwidth]{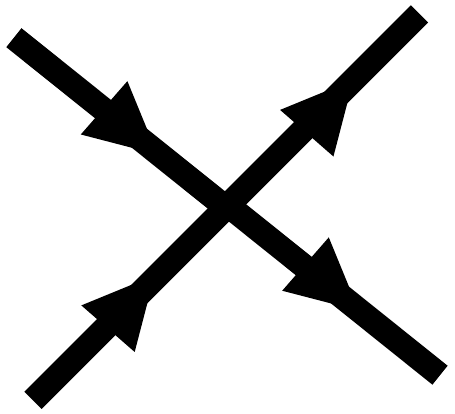}}&=-i\lambda_{RRAA}(k_{1},k_{2},k_{3},k_{4})\notag\\
&=
-i\lambda\Bigl[\bigl(1+f(k^{0}_{1})\bigr)\bigl(1+f(k^{0}_{2})\bigr)\notag\\
&\qquad\qquad-f(k^{0}_{1})f(k^{0}_{2})\Bigr]
\,,
\\
\parbox{2.cm}{\includegraphics[width=0.1\textwidth]{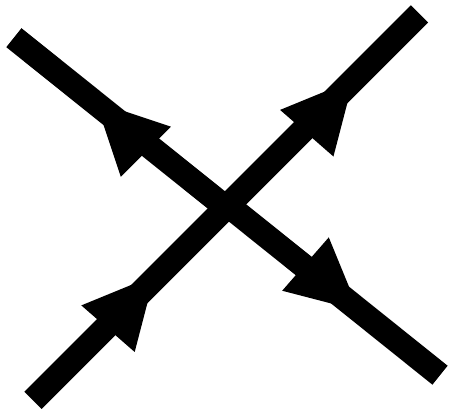}}&=
-i\lambda_{RAAA}(k_{1},k_{2},k_{3},k_{4})=
-i\lambda\,, \\ 
\parbox{2.cm}{\includegraphics[width=0.1\textwidth]{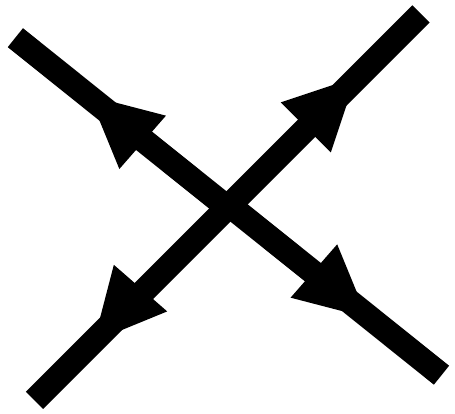}}&=
-i\lambda_{AAAA}(k_{1},k_{2},k_{3},k_{4})=0 \,.
\label{eq:feynmanRuleFourPoint}
\end{align}
At the tree level, vertices with all the same indices vanish, i.e, $\lambda_{RRRR}=\lambda_{AAAA}=0$.
This relation is satisfied in full $n$-point vertex functions; see Appendix~\ref{sec:identities}
for the derivation.
We should note, however, that this vanishment will be no longer  maintained for the system out of equilibrium,
where Kubo-Martin-Schwinger condition is not satisfied \cite{Gelis:2001xt}.

\begin{figure*}[t]
\includegraphics[width=0.7\textwidth]{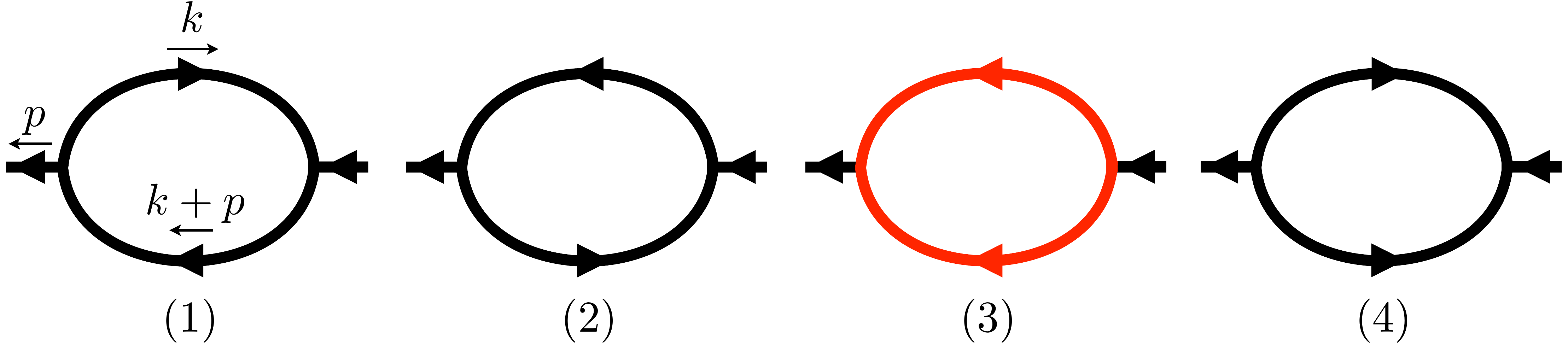}
\caption{One-loop diagram for a retarded Green function. The red line corresponds to the product of the retarded and advanced propagators.}
\label{fig:oneLoopDiagram}
\end{figure*}
\section{Computation method for Transport coefficients}
\label{sec:transportCoefficient}
In this section, we discuss the computational procedure for transport coefficients in the $R/A$ basis, and 
derive a self-consistent equation, which turns out to be an extension of the linearized Boltzmann equation.
We shall see that 
the $R/A$ basis can naturally lead to a decomposition of diagrams into the dominant diagrams 
including pinch poles and others.
\subsection{One-loop analysis} \label{oneLoopAnalysis} 

We shall start with the one-loop diagram of a retarded function.
Here, we employ the dressed propagators, $\DR(k)$ and $\DA(k)$, which do not have a pinch singularity. 
We assume that the relevant operator $\mathcal{O}(x)$ to the transport coefficient is a quadratic one.
The retarded Green function of it reads
\begin{equation}
\mathcal{G}^{\mathcal{O}}_{R}(p)= \int d^{4}x e^{ip\cdot x} i\theta(t)\langle 
[\mathcal{O}(x), \mathcal{O}(0)] \rangle \,.
\end{equation}
At one-loop level,
\begin{equation}
\begin{split}
\mathcal{G}^{\mathcal{O}}_{R}(p)&=\frac{i}{2}\int\frac{d^{4} k}{(2\pi)^{4}}
\bareThreePoint_{A\beta_1 \alpha_1 }(-p,-k,p+k)\\
&\quad\qquad\times D^{\alpha_{1}\alpha_{2}}(p+k)D^{\beta_{1}\beta_{2}}(-k) \\
&\qquad\qquad\times\bareThreePoint_{R\alpha_{2}\beta_{2}}(p,-p-k,k)\,,
\end{split}
\end{equation}
where the overall of $1/2$ is a symmetric factor, and the Feynman diagram for the vertex function, 
$\bareThreePoint_{\alpha\beta\gamma}(k_1,k_2,k_3)$ corresponding to $\mathcal{O}(x)$
is given by
\begin{align}
\parbox{2.cm}{\includegraphics[width=0.105\textwidth]{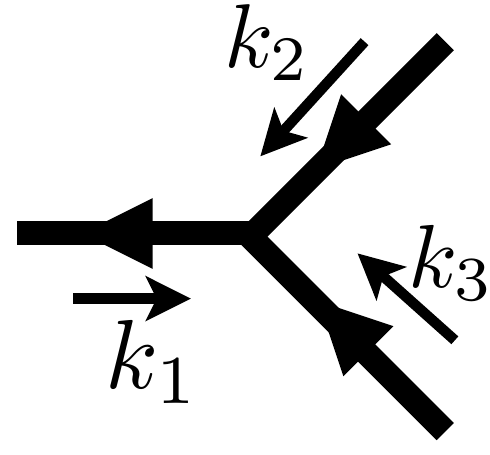}}&=\bareThreePoint_{ARR}(k_{1},k_{2},k_{3})
\notag\\
&=
 \bareThreePoint(k_{1},k_{2},k_{3})\bigl(1+f(k^0_2)+f(k^0_3)\bigr) \,, \\
 \parbox{2.cm}{\includegraphics[width=0.105\textwidth]{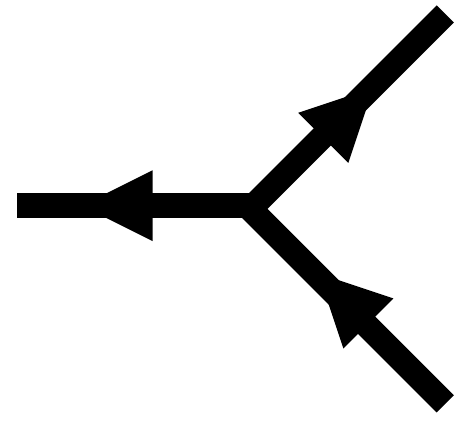}}&=
\bareThreePoint_{AAR}(k_{1},k_{2},k_{3})
=\bareThreePoint(k_{1},k_{2},k_{3}) \,,
\end{align}
and $\bareThreePoint_{RRR}(k_{1},k_{2},k_{3})=\bareThreePoint_{AAA}(k_{1},k_{2},k_{3})=0$.
For example, $\bareThreePoint(k_{1},k_{2},k_{3})=k_{2}^ik_{3}^{j}+k_{3}^ik_{2}^{j}-2\delta^{ij}\bm{k}_{2}\cdot\bm{k}_{3}/3$,
for the shear viscosity, where $\mathcal{O}(x)\propto \pi_{ij}(x)$.
There are, in general, four diagrams shown in Fig.~\ref{fig:oneLoopDiagram} contributing to the retarded Green function. 
The diagram $(4)$ vanishes because the vertex functions are 
$\bareThreePoint_{RRR}(k_{1},k_{2},k_{3})=\bareThreePoint_{AAA}(k_{1},k_{2},k_{3})=0$.
The diagram $(1)$ is expressed as 
\begin{equation}
\begin{split}
\mathcal{G}_R^{\mathcal{O}(1)}(p)&=\frac{i}{2}(-i)^2\int\frac{d^4k}{(2\pi)^4}\bigl(1+f(p^0)+f(k^0)\bigr)\\
&\qquad\times\varGamma^{(0)}(-p,-k,p+k)\DR(k)\DR(p+k)\\\
&\quad\qquad\times \varGamma^{(0)}(p,-p-k,k) \\
&=-\frac{i}{2}\int\frac{d^4k}{(2\pi)^4}f(k^0)\varGamma^{(0)}(-p,-k,p+k)\\
&\qquad\times\varGamma^{(0)}(p,-p-k,k)\DR(k)\DR(p+k) \,,
\end{split}
\end{equation}
where we have dropped the term proportional to $1+f(p^{0})$ that does not contain poles in the complex $k^0$ plane,
since the vertex and the retarded propagator have no poles in the upper complex $k^0$ plane; 
the $k^0$ integral becomes zero. 
For a transport coefficient, we need the soft-momentum limit as
\begin{equation}
\begin{split}
\transportCoefficient^{(1)}&\equiv\lim_{p^0\to0}\lim_{\bm{p}\to0}\frac{1}{p^0}\, \im \mathcal{G}_R^{\mathcal{O}(1)}(p)\\
&= \frac{1}{4}\int\frac{d^4k}{(2\pi)^4}\frac{\partial}{\partial k^0}f(k^0)\\
&\quad\qquad\times\bigl(\varGamma^{(0)}(0,k,-k)\bigr)^2 \;
\re\bigl( \DR(k) \bigr)^2 \,.
\end{split}
\end{equation}
For the contribution $\mathcal{G}_R^{\mathcal{O}(2)}(p)$ from the diagram (2), one finds $\transportCoefficient^{(2)}=\transportCoefficient^{(1)}$.
The contribution $\mathcal{G}_R^{\mathcal{O}(3)}(p)$ from the diagram (3) 
includes a product of the retarded and advanced propagators:
\begin{equation}
\begin{split}
\mathcal{G}_R^{\mathcal{O}(3)}(p)&=
\frac{i}{2}(-i)^2\int\frac{d^4k}{(2\pi)^4}\bigl(1+f(-k^0)+f(p^0+k^0)\bigr)\\
&\qquad\times\DA(k)\DR(p+k)
\varGamma^{(0)}(-p,-k,p+k)\\
&\qquad\qquad\times\varGamma^{(0)}(p,-p-k,k)
\\
&=-\frac{i}{2}\int\frac{d^4k}{(2\pi)^4}\bigl(f(p^0+k^0)-f(k^0)\bigr)\\
&\qquad\times\DA(k)\DR(p+k) 
\varGamma^{(0)}(-p,-k,p+k)\\
&\qquad\qquad\times\varGamma^{(0)}(p,-p-k,k) \,.
\end{split}
\end{equation}
Then, the transport coefficient
 $\transportCoefficient^{(3)}$ due to $\mathcal{G}_R^{\mathcal{O}(3)}(p)$
becomes
\begin{equation}
\begin{split}
\transportCoefficient^{(3)}
&\equiv\lim_{p^0\to0}\lim_{\bm{p}\to0}\frac{1}{p^0}\im \mathcal{G}_R^{\mathcal{O}(3)}(p)\\
&=-\frac{1}{2}\int\frac{d^4k}{(2\pi)^4}\frac{\partial}{\partial k^0}f(k^0)\\
&\qquad\qquad\times\frac{\rho(k)}{-2\im \varPi_R(k)} \bigl(\varGamma^{(0)}(0,k,-k)\bigr)^2\,.
\end{split}
\end{equation}

If one uses the quasiparticle approximation, $\rho(k)=2\pi \epsilon(k^0)\delta((k^0)^2-E_k^2)$ and $\gamma_k=-\im\varPi_R(E_k,\bm{k})/(2E_k)$,
$\transportCoefficient^{(3)}$ becomes
\begin{equation}
\begin{split}
\transportCoefficient^{(3)}&\simeq\frac{1}{T}\int \frac{d^3k}{(2\pi)^3}f(E_k)\bigl(1+f(E_k)\bigr)\\
&\quad\qquad\times
\frac{1}{2\gamma_k}
\left(\frac{\varGamma^{(0)}(0,k,-k)}{2E_k}\right)^2
\,,
\end{split}
\end{equation}
while $\transportCoefficient^{(1)}$ and  $\transportCoefficient^{(2)}$ vanish in this limit.
At weak coupling, the decay width is proportional to $\lambda^2$, so that $\transportCoefficient^{(3)}\sim 1/\lambda^2$.
This gives the correct coupling dependence, but not the coefficient. 
The higher-order diagram contributes to the transport coefficient
in the same order as the one-loop diagram.

Since the Bose-Einstein distribution function diverges at $k^0=0$, each $\mathcal{G}_R^{\mathcal{O}(i)}(p)$ includes divergence, although the sum of $\mathcal{G}_R^{\mathcal{O}(i)}(p)$'s is finite. In order to avoid such an artificial divergence,
we introduce an infrared (IR) cutoff $\IRCutoff$.
Bosons with very soft momentum can no longer be regarded as a particle,
so we have to treat it rather as wave.  This IR cutoff $\IRCutoff$ separates the scale between the particle and the wave.
Thus, we take the IR scale to be $\IRCutoff\ll m$. 
If one applies the quasiparticle approximation, the $\varPi_R(k)$ become independent of $\IRCutoff$, because $k^0\geq m\gg\kappa$ at the on-shell of the quasiparticle. 
We note that a fermion has no such a divergence because of the Fermi-Dirac distribution function.

The merit of the $R/A$ basis is that it enables us to identify 
 the diagrams with large contribution to $\mathcal{G}_R^{\mathcal{O}(3)}(p)$.
Although such an identification and decomposition are not obvious 
in the imaginary-time formalism before the analytical continuation,
the decomposition plays important role for the full analysis 
of the retarded Green function, as is discussed in the next subsection.

\subsection{Relativistic Eliashberg Decomposition  in Real-time Formalism}  \label{EliasbergMethod}
\begin{figure}
\includegraphics[width=0.45\textwidth]{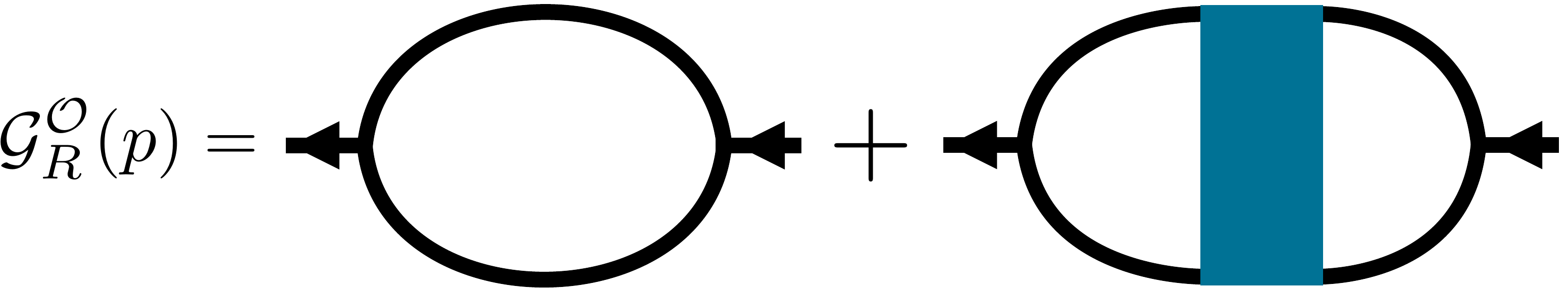}
\caption{A retarded Green function corresponding to a transport coefficient. The dark (green) square denotes the connected four-point function.
The Lines without arrows denote lines that can be the retarded or the advanced arrow lines.}
\label{fig:twoPointFunction}
\end{figure}
In the previous section, 
we showed within the one-loop level 
that the product of the retarded and advanced propagators with the 
common momentum gives a large contribution to transport coefficients.
Such a product of the propagators also appears in higher-order diagrams, which must be also 
summed over. 
Eliashberg method~\cite{Eliashberg:1962} is one of such a resummation method for non-relativistic
systems and has been applied to Fermi liquid at low temperature in the imaginary-time formalism.
In this method,  the four-point function is first analytically continued to the real-time domain.
 Each external leg can be the retarded or advanced legs,
so the four-point function has a matrix form. 
After the analytic continuation, the diagrams for the four-point function is decomposed into a set
that connects with pinching diagrams and  a set that does not connects with them. 
The pinching diagrams are summed over by a self-consistent equation, 
which is found to correspond to a kinetic equation.
In this section, we extend Eliashberg method to a relativistic case
using the real-time formalism  in the $R/A$ basis, and thus 
derive the corresponding relativistic kinetic equation.

The retarded Green function  shown diagrammatically in Fig.~\ref{fig:twoPointFunction} is expressed
in the momentum space as
\begin{widetext}
\begin{equation}
\begin{split}
\mathcal{G}^{\mathcal{O}}_{R}(p)&= \int d^{4}x e^{ip\cdot x} i\theta(t)\langle 
[\mathcal{O}(x), \mathcal{O}(0)] \rangle  \\
&=
\frac{i}{2}\int\frac{d^{4} k}{(2\pi)^{4}}\int \frac{d^{4} k'}{(2\pi)^{4}}
\bareThreePoint_{A\beta_1 \alpha_1 }(-p,-k,p+k)
D^{\alpha_{1}\alpha_{2}}(p+k)D^{\beta_{1}\beta_{2}}(-k) 
\Bigl[
(2\pi)^{4}\delta^{(4)}(k-k') 
{\delta_{\alpha_{2}}}^{\alpha_{4}}  {\delta_{\beta_{2}}}^{ \beta_{4} }\\
&\qquad\qquad+ (-i){\fourPointVertex}_{\alpha_{2}\beta_{2}\beta_{3}\alpha_{3}}(-p-k,k,-k',p+k') 
 D^{\alpha_{3}\alpha_{4}}(p+k')D^{\beta_{3}\beta_{4}}(-k')
 \Bigr]\bareThreePoint_{R\alpha_{4}\beta_{4}}(p,-p-k',k')\,,
 \end{split} 
 \label{eq:GR1}
 \end{equation}
 \end{widetext}
where  ${\fourPointVertex}_{\alpha_{2}\beta_{2}\beta_{3}\alpha_{3}}(-p-k,k,-k',p+k')$ denotes the connected 
four-point function.
The first diagram in RHS of Fig.~\ref{fig:twoPointFunction} 
is just the one discussed in the previous subsection.
Owing to Eq.~(\ref{eq:propagator}), the product of two propagators appearing in
Eq.~(\ref{eq:GR1}) is written as
 \begin{equation}
 \begin{split}
 &D^{\alpha_{1}\alpha_{2}}(p+k)D^{\beta_{1}\beta_{2}}(-k)\\
 &\quad=(-i)^{2}g^{\alpha_{1}\alpha_{2}}g^{\beta_{1}\beta_{2}}D^{\alpha_{1}}(p+k)D^{\beta_{1}}(-k)\\
&\quad \equiv -g^{\alpha_{1}\alpha_{2}}g^{\beta_{1}\beta_{2}}G^{\alpha_{1}\beta_{1}}(p,k) \,.
 \end{split}
 \label{eq:pairPropagators}
 \end{equation}
Thus we have
\begin{equation}
\begin{split}
\mathcal{G}^{\mathcal{O}}_{R}(p)&=
-\frac{i}{2}\int\frac{d^{4} k}{(2\pi)^{4}}\int \frac{d^{4} k'}{(2\pi)^{4}}
\bareThreePoint_{A\beta_1 \alpha_1 }(-p,-k,p+k) \\
&\quad\times G^{\alpha_{1}\beta_{1}}(p,k)
\Bigl[
(2\pi)^{4}\delta^{(4)}(k-k') 
{\delta^{\alpha_{1}}}_{\alpha_{2}}  {\delta^{\beta_{1}}}_{ \beta_{2} }
\\
&\qquad+ 
i{{\fourPointVertex}^{\alpha_{1}\beta_{1}}}_{\beta_{2}\alpha_{2}}(-p-k,k,-k',p+k')\\
&\qquad\qquad\times G^{\alpha_{2}\beta_{2}}(p,k')
 \Bigr]{\bareThreePoint_{R}}^{\alpha_{2}\beta_{2}}(p,-p-k',k')\,.
 \end{split} 
 \label{eq:GR2}
 \end{equation}
 The indices appear in pairs of $\alpha$ and $\beta$ in  Eq.~(\ref{eq:GR2}),
 so it is useful to introduce an index $l$ as  the pair of $R$ and/or $A$, i.e., $l=(\alpha,\beta)=(RA)$, $(AR)$, $(RR)$, and $(AA)$, then
\begin{equation}
\begin{split}
\mathcal{G}^{\mathcal{O}}_{R}(p)&=
-\frac{i}{2}\int\frac{d^{4} k}{(2\pi)^{4}}\int \frac{d^{4} k'}{(2\pi)^{4}}
\bareThreePoint_{A l}(p,k)
 G^{l}(p,k) \\
&\qquad\times \Bigl[
(2\pi)^{4}\delta^{(4)}(k-k') {\delta^{l}}_{m}\\
&\qquad\qquad+i{{\fourPointVertex}^{l}}_{m}(p,k,k')
G^{m}(p,k')
 \Bigr]{\bareThreePoint_{R}}^{m}(p,k') \,,
 \end{split} 
 \label{eq:GR3}
 \end{equation}
where the three-point functions  are 
\begin{align}
\;\;\parbox{2.cm}{\includegraphics[width=0.115\textwidth]{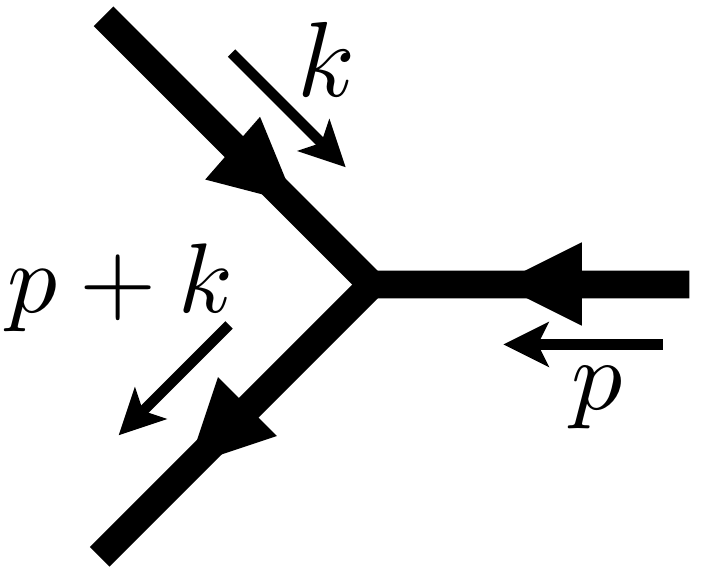}}&=
{\bareThreePoint_{R}}^{1}(p,k)  \notag\\
&= \bareThreePoint(p,-p-k,k)
\bigl(1+f(p^{0})+f(k^{0})\bigr)\,,
\end{align}
\begin{align}
\quad\parbox{2.cm}{\includegraphics[width=0.105\textwidth]{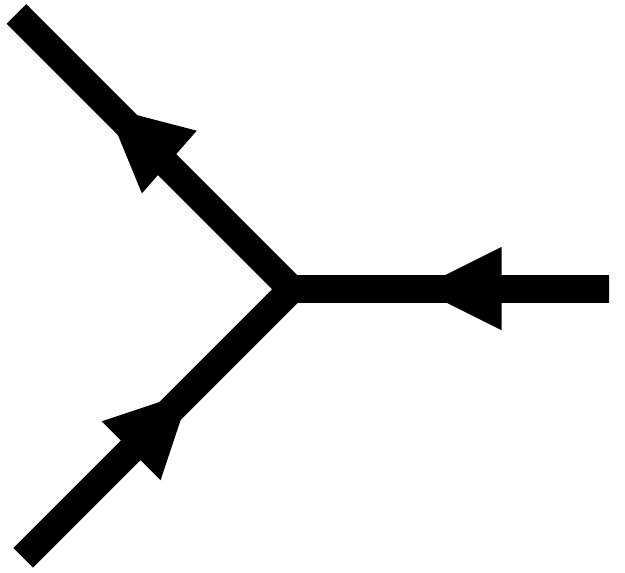}}&=
{\bareThreePoint_{R}}^{2}(p,k)
\notag\\
&= \bareThreePoint(p,-p-k,k)\notag\\
&\qquad\times\bigl(f(p^{0})-f(p^{0}+k^{0})\bigr)\,,\\
\quad\parbox{2.cm}{\includegraphics[width=0.105\textwidth]{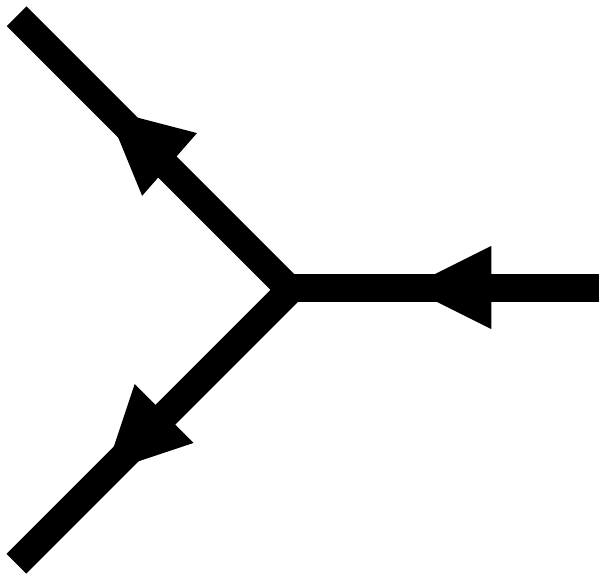}}&=
{\bareThreePoint_{R}}^{3}(p,k)
=\bareThreePoint(p,-p-k,k)\,,\\
\parbox{2.cm}{\includegraphics[width=0.115\textwidth]{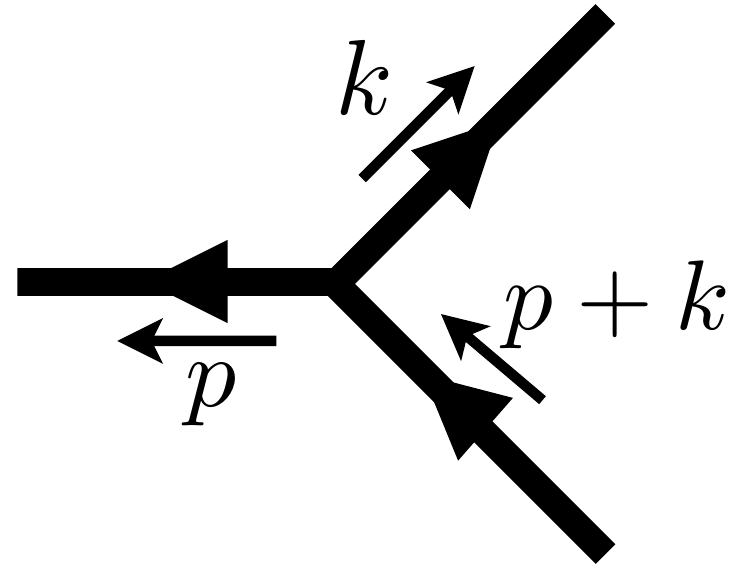}}&=
\bareThreePoint_{A 1}(p,k) 
 =\bareThreePoint(-p,-k,p+k)\,,\\
\quad\parbox{2.cm}{\includegraphics[width=0.105\textwidth]{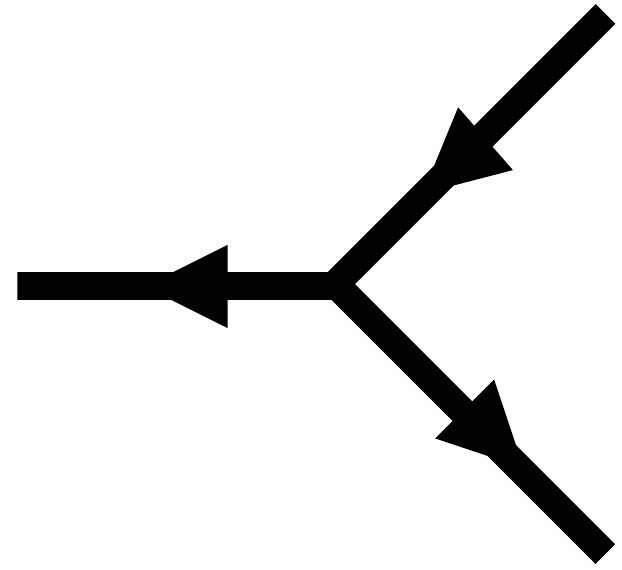}}
&=\bareThreePoint_{A 2}(p,k) 
= \bareThreePoint(-p,-k,p+k)\,,\\
\quad\parbox{2.cm}{\includegraphics[width=0.105\textwidth]{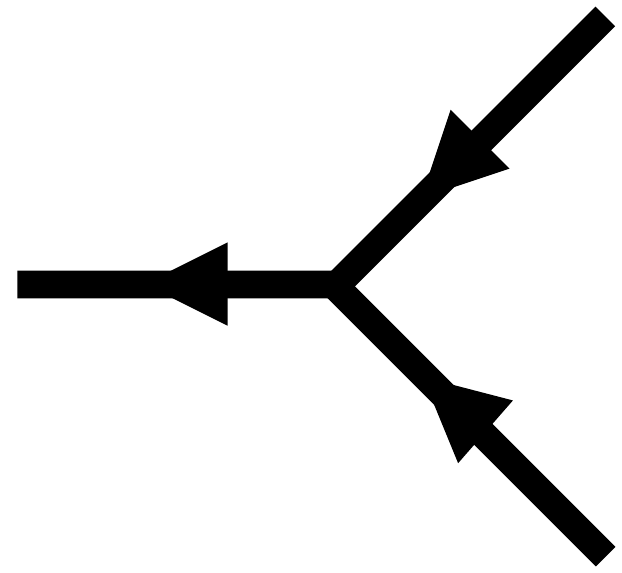}}&=
\bareThreePoint_{A 3}(p,k) 
\notag\\
&=\bareThreePoint(-p,-k,p+k)\notag\bigl(f(p^{0}+k^{0})-f(k^{0})\bigr)\,,
\end{align}
and $\bareThreePoint_{R 4}(p,k)=\bareThreePoint_{A 4}(p,k)=0$.
At the tree-level, the three-point function satisfies
\begin{equation}
\bareThreePoint_{A 3}(p,k) =\bigl(f(p^{0}+k^{0})-f(k^{0})\bigr)\bareThreePoint_{R 3}(-p,-k) \,.
\end{equation}
This relation is generalized to the full vertex function as
\begin{equation}
\varGamma_{A 3}(p,k) =\bigl(f(p^{0}+k^{0})-f(k^{0})\bigr)\varGamma^{*}_{R 3}(-p,-k) \,,
\end{equation}
the proof to which is given in Appendix~\ref{sec:identities}.

\begin{figure}
\includegraphics[width=0.48\textwidth]{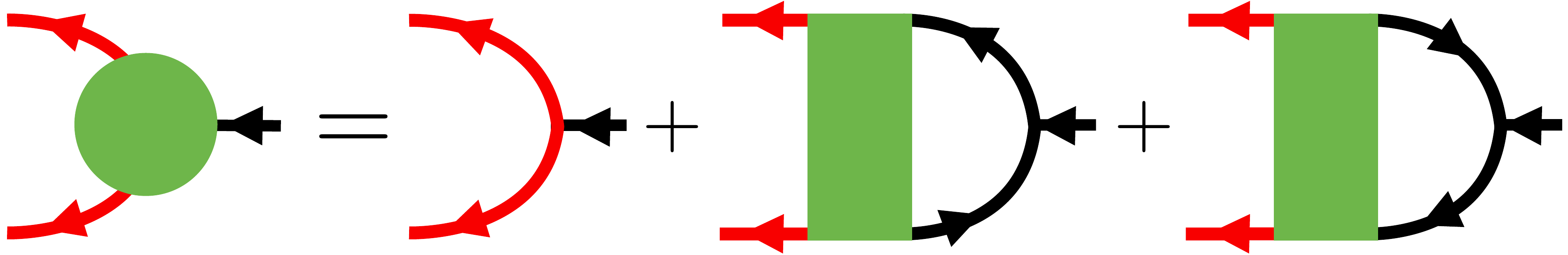}
\caption{Vertex corrections to $\bar{\varGamma}(p,k)$. The green square corresponds to ${{\fourPointVertexB}^3}_i(p,k,k')$.}
\label{eq:vertexRenormalization}
\end{figure}

As discussed in the previous section, $G^{3}(p,k)$ contains pinch singularities at weak coupling limit. 
Therefore we treat 
$G^{3}(p,k)$ separately from other $G^{i}(p,k)$ with $i\neq3$ in the four-point function.
Suppose that ${\fourPointVertexA}^{l}_{m}(p,k,k')$ is the four-point function that does not include 
a pair of lines of the type $G^{l}(p,k'')$.
\begin{figure}
\includegraphics[width=0.48\textwidth]{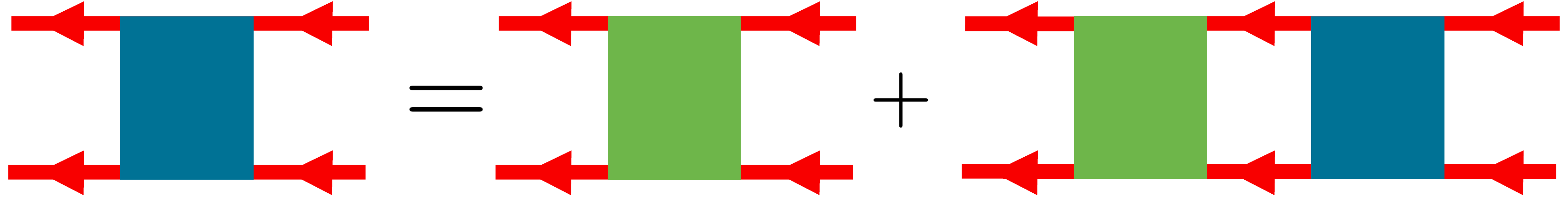}

\vspace{0.3cm}
\includegraphics[width=.48\textwidth]{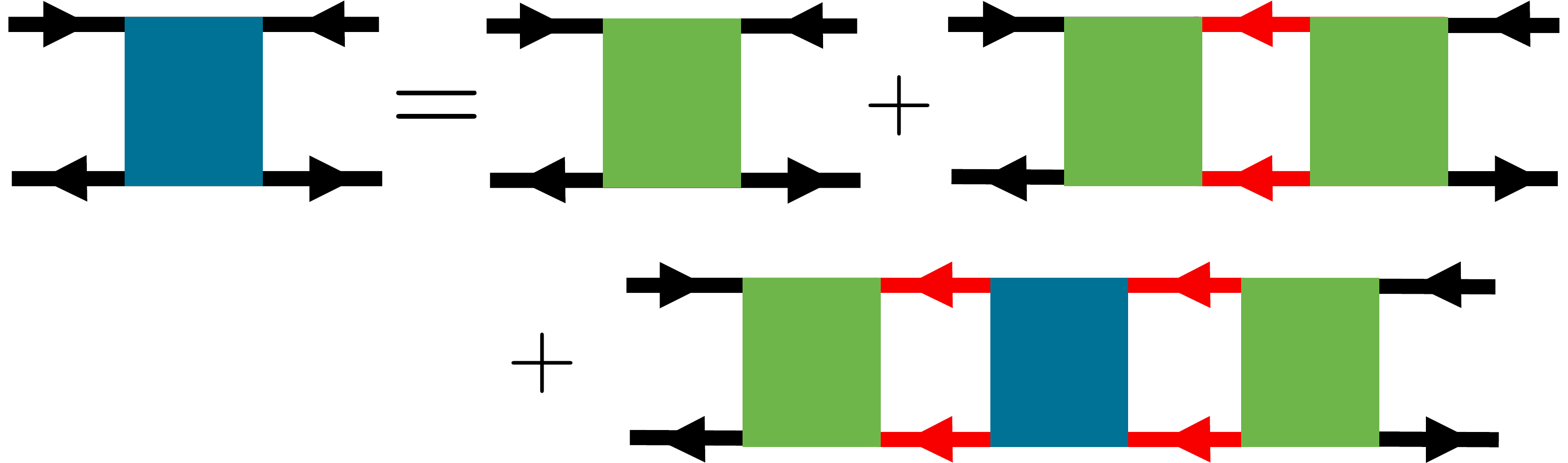}

\vspace{0.3cm}
\includegraphics[width=.48\textwidth]{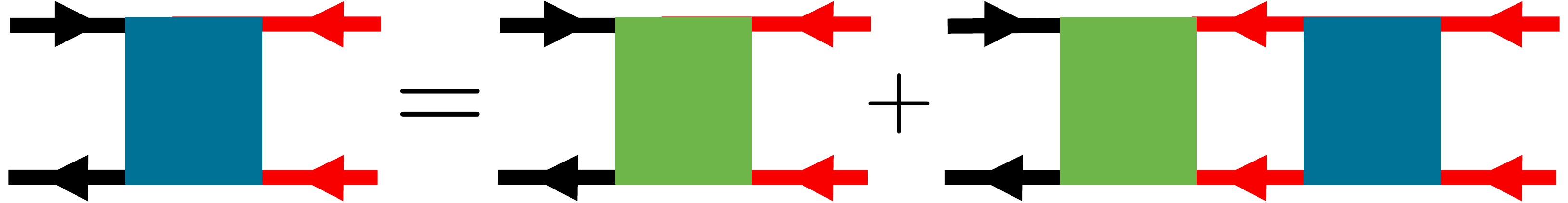}
\caption{Self-consistent equation for the four-point functions:
${\fourPointVertex^{3}}_3(p,k,k')$, ${\fourPointVertex^{1}}_2(p,k,k')$, and ${\fourPointVertex^{1}}_3(p,k,k')$, respectively.
The blue square denotes ${\fourPointVertex^{i}}_j(p,k,k')$ and the green square denotes ${{\fourPointVertexB}^{i}}_j(p,k,k')$.}
\label{fig:fourPointFunctions}
\end{figure}
Then, the four-point function obeys the following equation:
\begin{equation}
\begin{split}
{{\fourPointVertex}^{l}}_{m}(p,k,k')
&= {{\fourPointVertexA}^{l}}_{m}(p,k,k') \\
&\quad+(-i)^{3}\int \frac{d^{4}k''}{(2\pi)^{4}}
{{\fourPointVertexA}^{l}}_{n}(p,k,k'')\\
&\qquad\qquad\times G^{n}(p,k'')
{{\fourPointVertex}^{n}}_{m}(p,k'',k') \,.
\end{split}
\label{eq:selfConsistentEqForFourPointVertex}
\end{equation}
In order to pick $G^{3}(p,k)$ up, we also define ${{\fourPointVertexB}^l}_{m}(p,k,k')$ as a four-point function that does not include  $G^{3}(p+k'',k'')$.
The full four-point function ${{\fourPointVertex}^{3}}_{3}(p,k,k')$ obeys  the following self-consistent equation:
\begin{equation}
\begin{split}
{{\fourPointVertex}^{3}}_{3}(p,k,k')&={{\fourPointVertexB}^{3}}_{3}(p,k,k') \\
&\quad+(-i)^{3}\int \frac{d^{4}k''}{(2\pi)^{4}}
{{\fourPointVertexB}^{3}}_{3}(p,k,k'')\\
&\qquad\quad\times G^{3}(p,k'')
{{\fourPointVertex}^{3}}_{3}(p,k'',k') \,.
\end{split}
\label{eq:T33}
\end{equation}
Other full four-point functions satisfy
{\allowdisplaybreaks
\begin{align}
\begin{split}
{{\fourPointVertex}^{i}}_{j}(p,k,k')
&={{\fourPointVertexB}^{i}}_{j}(p,k,k') \\
&\quad+(-i)^{3}\int \frac{d^{4}k''}{(2\pi)^{4}}
{{\fourPointVertexB}^{i}}_{3}(p,k,k'')\\
&\qquad\qquad \times G^{3}(p,k'')
{{\fourPointVertexB}^{3}}_{j}(p,k'',k') \\
&\quad+(-i)^{6}\int \frac{d^{4}k''}{(2\pi)^{4}}\int \frac{d^{4}k'''}{(2\pi)^{4}}
{{\fourPointVertexB}^{i}}_{3}(p,k',k'')\\
&\qquad\qquad\times G^{3}(p,k'') 
{{\fourPointVertex}^{3}}_{3}(p,k'',k''')\\
&\qquad\qquad\qquad\times  G^{3}(p,k'''){{\fourPointVertexB}^{3}}_{j}(p,k''',k') \,,
\end{split}\\
\begin{split}
{{\fourPointVertex}^{i}}_{3}(p,k,k')
&={{\fourPointVertexB}^{i}}_{3}(p,k,k') \\
&\quad+(-i)^3\int \frac{d^{4}k''}{(2\pi)^{4}}
{{\fourPointVertexB}^{i}}_{3}(p,k,k'')\\
&\qquad\qquad\times G^{3}(p,k'')
{{\fourPointVertex}^{3}}_{3}(p,k'',k')  \,,
\end{split}
\\
\begin{split}
{{\fourPointVertex}^{3}}_{j}(p,k,k')
&={{\fourPointVertexB}^{3}}_{j}(p,k,k')\\
&\quad +(-i)^{3}\int \frac{d^{4}k''}{(2\pi)^{4}}
{{\fourPointVertex}^{3}}_{3}(p,k,k'')\\
&\qquad\qquad \times G^{3}(p,k'')
{{\fourPointVertexB}^{3}}_{j}(p,k'',k') \,.
\end{split}
\label{eq:T3j}
\end{align}
}
The diagrams corresponding to Eqs.~(\ref{eq:T33}) to (\ref{eq:T3j}) are shown 
in Fig.~\ref{fig:fourPointFunctions}.
Inserting Eqs.~(\ref{eq:T33}) to (\ref{eq:T3j}) into Eq.~(\ref{eq:GR3}), 
we arrive at a simple form:
\begin{equation}
\begin{split}
\mathcal{G}^{\mathcal{O}}_{R}(p)&=
\frac{i}{2} \int\frac{d^{4} k}{(2\pi)^{4}}\int \frac{d^{4} k'}{(2\pi)^{4}}
\bar{\varGamma}^{*}(-p,-k)\\
&\quad\times \bigl(f(k^{0}) - f(p^{0}+k^{0})\bigr)
G^{3}(p,k)\\
&\qquad\times\Bigl[
(2\pi)^{4}\delta^{(4)}(k-k')
+i{{\fourPointVertex}^{3}}_{3}(p,k,k')
G^{3}(p,k')
 \Bigr]\\
&\qquad \qquad\times\bar{\varGamma}(p,k') 
 +\mathcal{K}(p) \,,
\end{split}
\end{equation}
where 
\begin{equation}
\begin{split}
\mathcal{K}(p) &\equiv   
-\frac{i}{2}\sum_{i,j=1}^{2}\int\frac{d^{4} k}{(2\pi)^{4}}\int \frac{d^{4} k'}{(2\pi)^{4}}
\bareThreePoint_{A i}(p,k)
G^{i}(p,k) \\
&\qquad\times\Bigl[
(2\pi)^{4}\delta^{(4)}(k-k') {\delta^{i}}_{j}\\
&\qquad\qquad+i{{{\fourPointVertexB}}^{i}}_{j}(p,k,k')
G^{j}(p,k')
 \Bigr]{\bareThreePoint_{R}}^{j}(p,k') \,,
\end{split}
\end{equation}
which does not include $G^{3}(p,k)$, so that $\mathcal{K}(p)$ is not enhanced at small $p$.
We have also introduced the effective-vertex function shown diagrammatically 
in Fig.~\ref{eq:vertexRenormalization},
which is expressed as
\begin{equation}
\begin{split}
\bar{\varGamma}(p,k)&\equiv
\bareThreePoint(p,k)\\
&\quad+(-i)^{3}\sum_{i=1}^{2}\int\frac{d^{4}k'}{(2\pi)^{4}}
{{\fourPointVertexB}^{3}}_{i}(p,k,k')\\
&\qquad\qquad\qquad\times G^{i}(p,k')
{\bareThreePoint_{R}}^{i}(p,k') 
\,.
\label{eq:vertexRen}
\end{split}
\end{equation}
This effective-vertex function includes vertex corrections which can be evaluated by loop expansions 
as long as further infrared singularities do not appear,
and contains both quantum and medium effects;
these corrections are in the same order in coupling at high temperature.
We emphasize that this vertex correction is not included in the usual kinetic theory, 
and hence the vertex correction derived here describes an effect beyond the kinetic theory.

Now the transport coefficient $\transportCoefficient$ is given 
by the retarded Green function at the $p\to0$ limit, 
\begin{equation}
\transportCoefficient \equiv\lim_{p^{0}\to0}\lim_{\bm{p}\to\bm{0}}
 \frac{1}{p^{0}}\, \im\mathcal{G}_{R}(p)\equiv \transportCoefficient_{B}+\transportCoefficient_{\mathcal{K}}\,.
\label{eq:transportCoefficient}
\end{equation}
Note that these limits, $p^0\to0$ and $\bm{p}\to\bm{0}$, do not exchange in general.
We have decomposed the transport coefficient into two parts,
$\transportCoefficient_{B}+\transportCoefficient_{\mathcal{K}}$, the explicit forms of which
will be given shortly.

Let us discuss the two terms one by one.
The first term is expressed as 
\begin{equation}
\begin{split}
\transportCoefficient_{B}&\equiv\frac{1}{2}\int\frac{d^{4} k}{(2\pi)^{4}}\int \frac{d^{4} k'}{(2\pi)^{4}}
\bar{\varGamma}^*(-k) \left(-\frac{\partial}{\partial k^{0}}f(k^{0})\right)\\
&\qquad\times\frac{\rho(k)}{-2\,\im\varPi_{R}(k)} 
\Bigl[
(2\pi)^{4}\delta^{(4)}(k-k')\\
&\qquad\qquad-\im{{\fourPointVertex}^{3}}_{3}(k,k')
\frac{\rho(k')}{-2\,\im\varPi_{R}(k')}
 \Bigr]\bar{\varGamma}(k')  \,,
\end{split} 
\label{eq:transportCoefficent}
\end{equation}
where we have used the relation
\begin{equation}
G^{3}(0,k)=\DR(k)\DA(k)=\frac{\rho(k)}{-2\,\im\varPi_{R}(k)} \,,
\end{equation}
in accordance with Eq.~(\ref{eq:pairPropagatorAtP=0}).
We have also introduced abbreviated notations:
\begin{align}
{{\fourPointVertex}^{3}}_{3}(k,k') &=  {{\fourPointVertex}^{3}}_{3}(0,k,k')\,,  \\
 \bar{\varGamma}(k') &= \bar{\varGamma}(0,k') \,, \\
\bar{\varGamma}^*(k) &= \bar{\varGamma}^{*}(0,k)\,.
\end{align}
The second term of RHS in Eq.~(\ref{eq:transportCoefficent}) is  defined by 
\begin{equation}
\transportCoefficient_{\mathcal{K}}\equiv\lim_{p^{0}\to0}\lim_{\bm{p}\to\bm{0}}\frac{1}{p^{0}}\,\im\mathcal{K}(p) \,.
\label{eq:sigma_kappa}
\end{equation}

As will be clarified in the next subsection, 
the physics content of the $\transportCoefficient_{B}$ can be
nicely given  in terms of  a Boltzmann equation.
At weak coupling, $\transportCoefficient_{B}$ becomes much larger 
than $\transportCoefficient_{\mathcal{K}}$ because of the pinch singularity.
In the following, we focus on $\transportCoefficient_{B}$.

\begin{figure}
\includegraphics[width=0.48\textwidth]{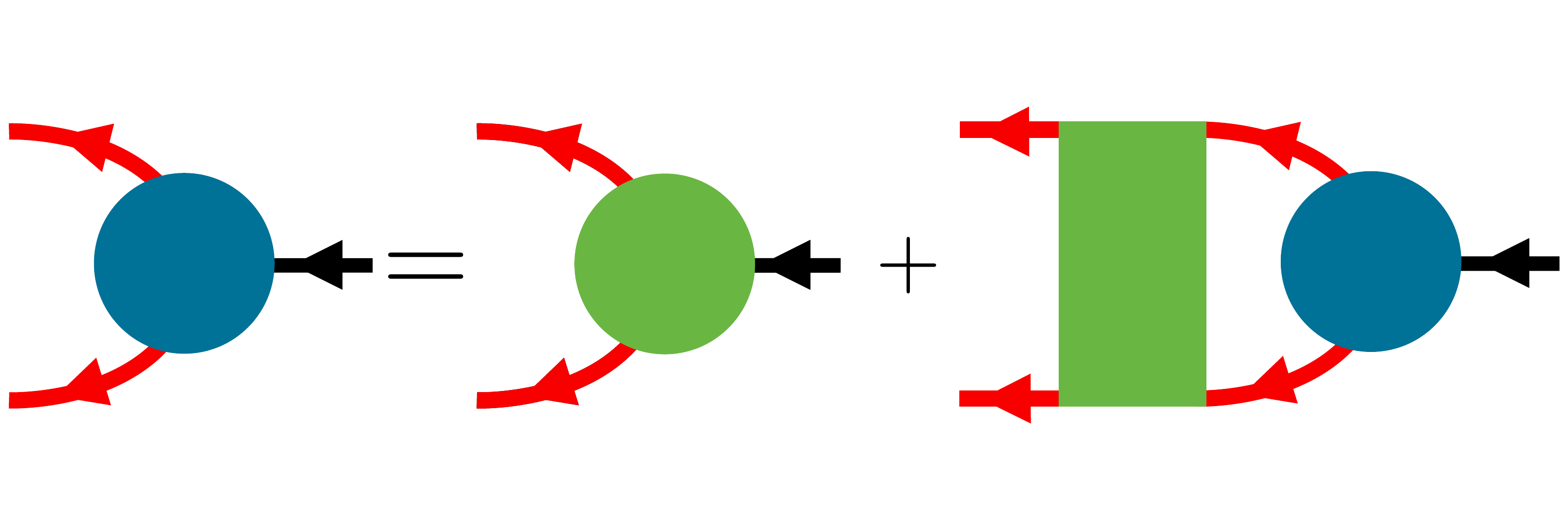}
\caption{The self-consistent equation for the vertex function. The blue and green blob corresponds to the full and renormalized vertex function.
The green square corresponds to the collision term.}
\label{fig:selfconsistentEquation}
\end{figure}

Let us introduce the full vertex function, $\varGamma(k)$, by
\begin{equation}
\begin{split}
\varGamma(k)&=\bar{\varGamma}(k)\\
&\quad- \int \frac{d^{4} k'}{(2\pi)^{4}}\im{{\fourPointVertex}^{3}}_{3}(k,k')
\frac{\rho(k')}{-2\im\varPi_{R}(k')}
\bar{\varGamma}(k') \,,
\end{split}
\end{equation}
which is found to obey the following self-consistent equation on account of 
Eq.~(\ref{eq:T33}):
\begin{equation}
\begin{split}
\varGamma(k)&=\bar{\varGamma}(k)\\
&\quad-\int \frac{d^{4}k'}{(2\pi)^{4}}
\im{{\fourPointVertexB}^{3}}_{3}(k,k')
\frac{\rho(k')}{-2\im\varPi_{R}(k')}
\varGamma(k')\,.
\end{split}
\label{eq:BetheSalpeterEquation}
\end{equation}
The diagrammatical representation of this integral equation
is given in Fig.~\ref{fig:selfconsistentEquation}.

Now ${{\fourPointVertexB}^{3}}_{3}(p,k,k')$ consists of two parts: 
One is ${{\fourPointVertexA}^{3}}_{3}(p,k,k')$ that does not include $G^{i}(p,k)$,
and the other $G^{i}(p,k)$. 
The latter part contains ${{\fourPointVertexA}^{i}}_{3}(p,k,k')$, 
which is proportional to $f(p^{0}+k'^{0})-f(k'^{0})$, 
so that it vanishes as $p^{0}\to0$. 
Therefore, ${{\fourPointVertexB}^{3}}_{3}(k,k')={{\fourPointVertexA}^{3}}_{3}(k,k')$.
The dominant transport coefficient $\transportCoefficient_{B}$ is now expressed as
\begin{equation}
\begin{split}
\transportCoefficient_{B}=&
 \frac{1}{2T}\int\frac{d^{4} k}{(2\pi)^{4}}\frac{\rho(k)}{-2\im\varPi_{R}(k)}\\
&\qquad\times f(k^{0})\bigl(1+f(k^{0})\bigr)\bar{\varGamma}^{*}(k)
\varGamma(k) \,,
\end{split}
\label{eq:transportCoefficient2}
\end{equation}
where we have used the relation $\bar{\varGamma}^{*}(k)=\bar{\varGamma}^{*}(-k)$.

Equations~(\ref{eq:vertexRen}), (\ref{eq:BetheSalpeterEquation}), and (\ref{eq:transportCoefficient2}) are our main result.
What we have done is rewriting diagrams in a useful and physical form. 
This is just a rewriting of diagrams, so that there is no approximation in the sense of the diagrammatic expansion
(although nonperturbative contributions such as instantons are not included.)
The $\varPi_R(k)$, ${{\fourPointVertexA}^3}_3(k,k')$, and $\bar{\varGamma}(k)$ can be expanded by loops as far as further infrared singularities appear. The spectral function is obtained through  $\varPi_R(k)$.
One has to choose  the loop diagrams for $\varPi_R(k)$ and ${{\fourPointVertexA}^3}_3(k,k')$ so as to satisfy the symmetries of the action,
 i.e., Ward-Takahashi identity, as is discussed in  Appendix~\ref{sec:WTIdentity}.

Let us estimate the order of $\transportCoefficient_B$ at weak coupling.
The leading contribution comes from the peak of the spectral function.
The residue of the pole corresponding to the peak is of order 1, and 
$\im\varPi_R(k)$ is of order $\lambda^2$ at the pole, 
so the first part in Eq.~(\ref{eq:transportCoefficient2}) gives a contribution of order $1/\lambda^2$.
The four-point function, $\im{{\fourPointVertexA}^{3}}_{3}(k,k')$, in Eq.~(\ref{eq:BetheSalpeterEquation}) is related to the squared scattering-amplitude,
which is of order $\lambda^2$ from $2\to2$ scattering. This is the same order as the imaginary part of the self-energy,
so the full vertex $\varGamma(k)$ is the same order as $\bar{\varGamma}(k)$. 
As we will see in the next subsection, $\im{{\fourPointVertexA}^{3}}_{3}(k,k')$ 
and $\im\varPi_{R}(k)$ are related to the collision term of a Boltzmann equation, 
which has the same order of the coupling constant.
As a result, the $\transportCoefficient_B$ is estimated as of order $1/\lambda^2$ at weak coupling.
This is consistent with the result in kinetic theory where the shear viscosity is proportional to the inverse of the transport cross-section of order $\lambda^2$.
We note that the $\transportCoefficient_B$ diverges at the zero cutoff limit, $\kappa=0$, for bosons, because $f(k)$ has the pole at $k^0=0$,
while fermions do not. The divergence will be cancelled by adding $\transportCoefficient_\mathcal{K}$.

\subsection{Linearized Boltzmann Equation} \label{sec:LinearizedBoltzmannEquation}

Here we discuss the relation between our diagrammatic method and the linearized Boltzmann equation.
We will show that Eq.~(\ref{eq:BetheSalpeterEquation})  has the form of a linearized Boltzmann equation.
In the leading order of the coupling constant, 
it is known that the linear equation, Eq.~(\ref{eq:BetheSalpeterEquation}), with the quasiparticle approximation is reduced to a linearized Boltzmann equation in scalar theory \cite{Jeon} and QED \cite{Gagnon}.
Beyond the leading order, nothing has been known about the relation between diagrammatic method for Kubo formula and kinetic equation.
Although the Boltzmann equation is the equation for the on-shell (quasi-) particles,
the spectral function includes not only the quasiparticle peak but also multi-particles state spectrum.
If one wants to derive the Boltzmann-like equation, which contains the scattering process of the quasiparticles, 
one needs to decompose the spectrum function into the quasiparticle part with a distinct peak
 and others, and rewrite the equation in terms of the quasiparticles. 
In this paper, we shall not do such a decomposition; nevertheless, we shall show the linear equation has the similar property to that of linearized Boltzmann equation.
To derive a linearized collision operator in our formalism, we define $\varphi(k)$ by
\begin{equation}
\varphi(k) \equiv \frac{\varGamma(k)}{-2\,\im\varPi_{R}(k)} \,,
\label{eq:phi}
\end{equation}
following \cite{Jeon}.
Then, Eq.~(\ref{eq:BetheSalpeterEquation}) is cast into the following form
\begin{equation}
\frac{1}{2k^0}\bar{\varGamma}(k)= \mathcal{L}\,\varphi(k) \,,
\label{eq:linearizedBoltzmannEquation}
\end{equation}
where 
$\mathcal{L}$ is defined by
\begin{equation}
\begin{split}
\mathcal{L}\,\varphi(k) &\equiv
-2\,\frac{1}{2k^0}\,\im\varPi_{R}(k)\varphi(k)\\
&\quad+\frac{1}{2k^0}\int \frac{d^{4}k'}{(2\pi)^{4}}\rho(k')\,
\im{{\fourPointVertexA}^{3}}_{3}(k,k')
\varphi(k') \, .
\end{split}
\label{eq:collisionOperator}
\end{equation}
We note that $\varphi(k)$ is an odd (even) function under $k\leftrightarrow-k$, if $\varGamma(k)$ is an even (odd) function,
 since $\im\varPi_R(k)$ is an odd function. 
Equation~(\ref{eq:linearizedBoltzmannEquation}) has the same form 
as the linearized Boltzmann equation~\cite{linearizedBoltzmann}.
In fact, $\mathcal{L}$ can  be identified with the linearized collision operator:
The equivalence between Eq.~(\ref{eq:linearizedBoltzmannEquation}) and the linearized Boltzmann equation
in the leading order of the coupling constant is shown in Appendix~\ref{sec:BoltzmannVSKuboFormula}.

Here we introduce an inner product 
as
\begin{equation}
\langle\varphi_{1},\varphi_{2}\rangle \equiv \int \frac{d^{4}k}{(2\pi)^{4}}W(k) \varphi_{1}^{*}(k) \varphi_{2}(k) \,,
\label{eq:innerProduct}
\end{equation}
where $W(k)\equiv k^0(1+f(k^{0}) )f(k^{0}) \rho(k) >0$ is the weight function, and $\varphi_1$ and $\varphi_2$ are arbitrary functions of $k^\mu$.
We note  that $W(k)$ at $k_0=0$ is finite, although $f(k^{0})$ has a singularity at $k_0=0$. 
Using this inner product, we can rewrite the transport coefficient, Eq.~(\ref{eq:transportCoefficient2}) as
\begin{equation}
\transportCoefficient_{B}=
\frac{1}{T} \langle \mathcal{S}, \mathcal{L}^{-1} \mathcal{S}\rangle \,,
\label{eq:transportCoefficeintBoltzmann}
\end{equation}
where $\mathcal{S}\equiv\bar{\varGamma}/(2k^0)$.
Note that $\mathcal{S}$ must be orthogonal to the eigenvectors of $\mathcal{L}$
with zero eigenvalues corresponding to conserved charges;
otherwise $\mathcal{L}^{-1}$ is not well-defined.

The inverse of the collision operator can be expressed as
\begin{equation}
\mathcal{L}^{-1}=\int_0^\infty dt e^{-t \mathcal{L}} \,,
\label{eq:timeEvolution}
\end{equation}
where we have assumed that the eigenvalues of $\mathcal{L}$ are nonnegative.
This is necessary for the stability of the system.
Using Eq.~(\ref{eq:timeEvolution}), Eq.~(\ref{eq:transportCoefficeintBoltzmann}) is rewritten to
a Kubo formula for the quasiparticle,
\begin{equation}
\transportCoefficient_{B}=
\frac{1}{T}\int_0^\infty dt
\langle\mathcal{S}(0), \mathcal{S}(t) \rangle \,,
\label{eq:KuboFormulaBoltzmann}
\end{equation}
where $\mathcal{S}(t)=\exp(-t\mathcal{L})\mathcal{S}(0)$.
We rewrite a Kubo formula in field theory to that for relevant quasiparticles.
Equation~(\ref{eq:KuboFormulaBoltzmann}), a semiclassical formula,
i.e.,  no field operator is present in Eq.~(\ref{eq:KuboFormulaBoltzmann}).
The collision operator and the effective vertex are calculated in thermal field theory.
We note that the thermal weight $W(k)$ does not contain $f(k^0)$ but $f(k^0)(1+f(k^0))$,
because the thermal average is taken for a two-point correlation function.
If the $\mathcal{S}$ is an eigenstate of $\mathcal{L}$ with an eigenvalue, $1/\tau$,
\begin{equation}
\transportCoefficient_{B}=
\frac{\tau}{T} \langle  \mathcal{S}(0),\mathcal{S}(0)\rangle \,.
\label{eq:transportCoefficientClassical}
\end{equation}
The inverse of the eigenvalue $\tau$ can be identified as a relaxation time.

Let us estimate the shear viscosity using Eq.~(\ref{eq:transportCoefficientClassical}) in the quasiparticle limit, 
where, the thermal weight function becomes
\begin{equation}
W(k)\simeq2\pi\bigl[\delta(k^0-|\bm{k}|)+ \delta(k^0+|\bm{k}|)\bigr]f(|\bm{k}|)\bigl(1+f(|\bm{k}|)\bigr) \,,
\end{equation}
and the vertex function for the shear viscosity $\pi^{ij}$ in the leading order is
\begin{equation}
\pi^{ij}= \frac{1}{k^0}\left(k^ik^j-\delta^{ij}\frac{\bm{k}^2 }{3}\right) \,.
\label{eq:vertexFunctionShearViscosity}
\end{equation}
Then, the shear viscosity is evaluated as
\begin{equation}
\eta =\frac{\tau}{10T}\langle \pi^{ij},\pi_{ij}\rangle=\frac{\tau sT }{5} \,.
\label{eq:viscosityRelaxationtime}
\end{equation}
where $s=2\pi^2T^3/45$ is the entropy density in the free theory.
Using $P=4s$ for free massless particles in the leading order, we obtain the relaxation time as
\begin{equation}
\tau = \frac{5}{4}\frac{\eta}{P} \,.
\end{equation}
This is $25\%$ larger than that obtained in Ref.~\cite{Koide:2009sy},
although Eq.~(\ref{eq:viscosityRelaxationtime}) is a parametrically similar to that given
 in Ref.~\cite{Koide:2009sy}.
The difference comes from the semi-classical calculation in our formalism
and the field theoretical calculation in Ref.~\cite{Koide:2009sy}.

\subsection{Leading and higher orders}

In this subsection we show how the contributions of the higher
 as well as  leading-order terms to the transport coefficient appear in our formalism. 
The transport coefficient $\transportCoefficient_{B}$ in Eq.~(\ref{eq:transportCoefficeintBoltzmann})
 consists of the vertex function $\mathcal{S}$, the collision term $\mathcal{L}$ 
and the inner product that contains the spectral function $\rho(k)$.
The contribution of the leading and higher order terms are nicely  summarized into these terms.
The merit of our formalism is that the contributions of 
each term can be systematically estimated using loop expansions.
To show the details of this statement, let us consider the shear viscosity as an example.
We will find the shear viscosity is expanded by the coupling constant $\lambda$ as
\begin{equation}
\frac{\eta}{\eta_0}= 1+ c_1\sqrt{\lambda}+(c_2'\ln1/\lambda+c_2)\lambda + \cdots \,.
\label{eq:eta/eta_0}
\end{equation}
Here the numerical calculation in the leading-order gives
$\eta_0 \simeq 3033.54  T^3/\lambda^2$~\cite{Jeon,nextLeading1}.

\begin{figure}
\begin{center}
\includegraphics[width=0.3\textwidth]{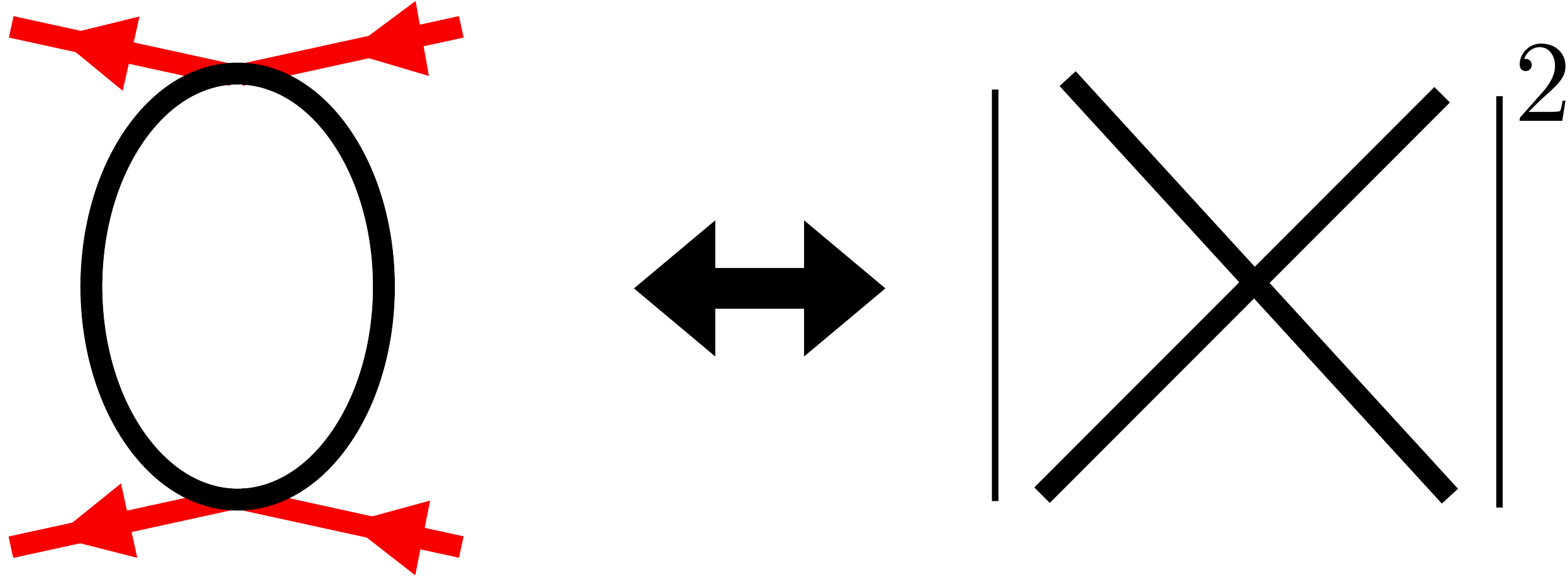}
\end{center}
\caption{The diagram contributing to ${{\fourPointVertexA}^3}_3(k,k')$ in the leading order, which corresponds to $2\to2$ scattering.}
\label{fig:collisionTermLeading}
\end{figure}

First, let us begin with clarifying how the $\lambda^{-2}$-dependence
 of 
$\eta_0$ is obtained in the leading order within our formalism,
although a biref discussion on this matter was given in 
the end of subsection~\ref{EliasbergMethod}.
The diagram of ${{\fourPointVertexA}^3}_3(k,k')$ in the leading order is shown in 
Fig.~\ref{fig:collisionTermLeading} that corresponds to the $2\to2$ scattering.
Then the squared scattering amplitude is proportional to $\lambda^2$,
and thus the collision term is estimated as 
$\mathcal{L}\sim{{\fourPointVertexA}^3}_3(k,k')\sim\lambda^2$.
In this calculation, the quasiparticle approximation with a vanishing mass is employed
in the spectral function, $\rho(k)=(2\pi)\text{sgn}(k^0)\delta(k^2)$,
because hard momenta of order $k\sim T$ contribute to the leading order, where
the mass can be neglected.
Formally, 
the diagram in LHS of Fig.~\ref{fig:collisionTermLeading} also contains $1\to3$ decay and 
$3\to1$ fusion processes; however, they are found to be higher orders of the coupling 
because of the phase space restricted by the enegy-momentum conservation.
The contribution of $\mathcal{S}$ is of order one.
As a result, we have $\eta_0\sim \mathcal{S}^2/\mathcal{L}\sim1/\lambda^2$,
which coincides with the result given by 
the Boltzmann equation. 
In fact, the equivalence  between the diagrammatic method in the leading order 
and the linearized Boltzmann equation generally holds, 
as is proved in appendix~\ref{sec:BoltzmannVSKuboFormula}.

Next, let us estimate the contribution of the next leading order (NLO) to 
the shear viscosity in Eq.~(\ref{eq:eta/eta_0}), 
which contributions are found to come from the correction to $\rho(k)$,
while the loop corrections to $\mathcal{S}$ and $\mathcal{L}$, at least of order $\lambda$,
are higher-order corrections.
The spectral function are characterized by
the thermal mass, width and height of the peak of the quasiparticle.
In the NLO, the thermal-mass correction gives the contribution of order $\sqrt{\lambda}$,
which is nonanalytic in $\lambda$.~\cite{nextLeading1}.
The nonanalytic term  is obtained from a phase space integral with
an infrared enhancement of the Bose-Einstein distribution function $f(E)\simeq T/E$ 
at small $E$.
Since $\rho(k)$ appears with $f(k)$ in $\mathcal{L}$, $\mathcal{S}$, and the inner product (see, e.g., Eq.~(\ref{eq:collisionTemLeading}),)
consider the following integral to see the nonanalyticity of $\lambda$:
\begin{equation}
\begin{split}
&\int\frac{d^4p}{(2\pi)^4}\rho(p)f(p)\\&= \frac{T^2}{12}\Bigl[1+ a_1 \frac{m_T}{2\pi T}
 +\!\Bigl(a_2'\ln \Bigl(\frac{m_T}{2\pi T}\Bigr)^2\!\!+a_2\Bigr) \Bigl(\frac{m_T}{2\pi T}\Bigr)^2\!\!+\cdots\Bigr] \,,
\end{split}
\label{eq:phaseSpaceIntegral}
\end{equation}
where $a_1=6$, $a_2=3-6\gamma_E+6\ln 2$ and $a'_2=3$ with the Euler constant $\gamma_E=0.577216$.
We have assumed that the spectral function has a form 
$\rho(k)=(2\pi)\text{sgn}(k^0)\delta(k^2-m_T^2)$ to obtain Eq.~(\ref{eq:phaseSpaceIntegral}).
The nonanalytic term $m_T$ appears in RHS of Eq.~(\ref{eq:phaseSpaceIntegral}),
although the integrant is a function of $m_T^2$, which is obtained as~\cite{Altherr:1989rk}
\begin{equation}
m_T^2=\frac{\lambda T^2}{24}\left[ 1-3\left(\frac{\lambda}{24\pi^2}\right)^{\frac{1}{2}}+\cdots\right] \,.
\label{eq:thermalMass}
\end{equation}
The thermal masses in the leading ($\sim\lambda$) and the next leading  ($\sim\lambda^{3/2}$) orders 
are given by the one-loop diagram and the ring diagram resummation, respectively.
Therefore the contribution of the phase space integral in the NLO gives that of order $\sqrt{\lambda}$, and the coefficient is estimated as $c_1\simeq 0.104$~\cite{nextLeading1}.
We note that this term also contains the next-to-next-leading order (NNLO) corrections of orders $\lambda\ln1/\lambda$ and $\lambda$ to the shear viscosity from Eqs.~(\ref{eq:phaseSpaceIntegral}) and (\ref{eq:thermalMass}).
One might worry about corrections from the thermal width of the spectral function;
however, this is in even higher-orders
because the width is of order $\lambda^2 T$, which is smaller than $\lambda T$.

\begin{figure}
\begin{center}
\includegraphics[width=0.47\textwidth]{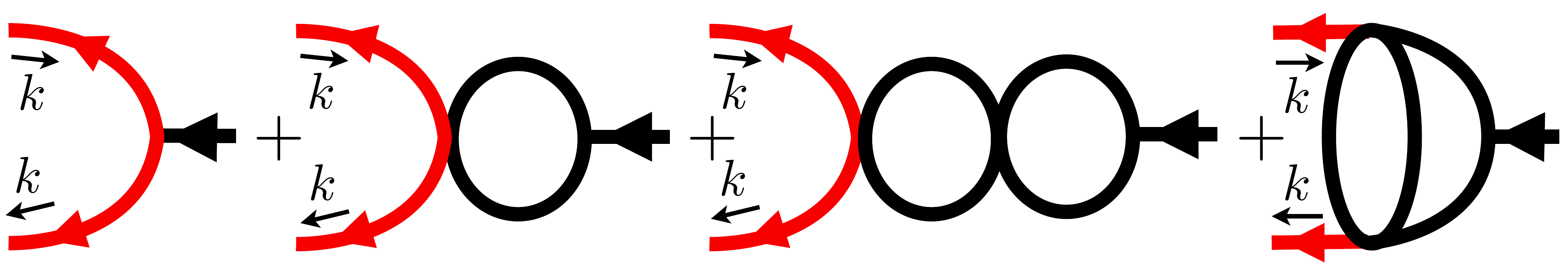}
\end{center}
\caption{Vertex corrections up to of order $\lambda^2$.}
\label{fig:vertexCorrections}
\end{figure}
\begin{figure}
\begin{center}
\includegraphics[width=0.47\textwidth]{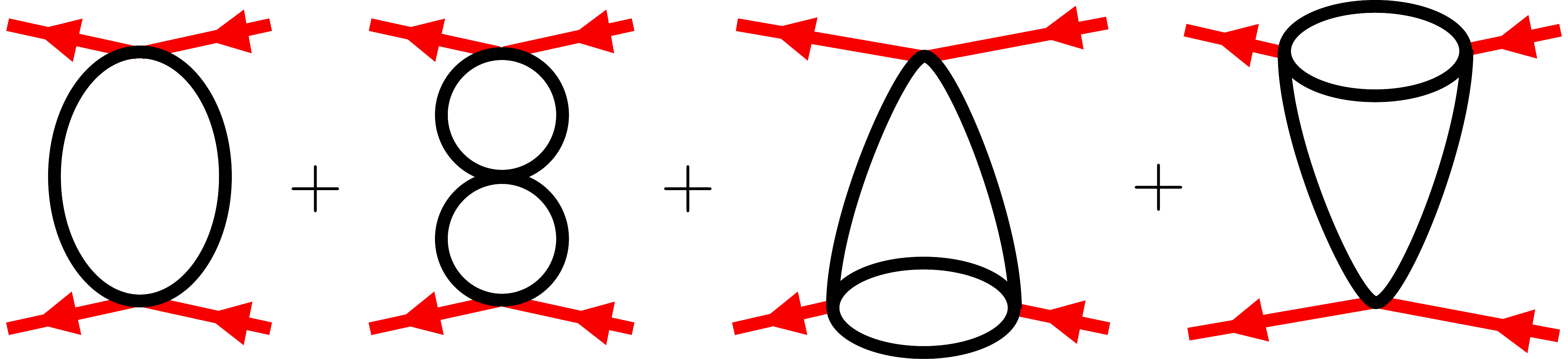}
\end{center}
\caption{Diagrams of order $\lambda^3$  contributing to ${{\fourPointVertexA}^3}_3(k,k')$.}
\label{fig:collisionTerm}
\end{figure}
\begin{figure}
\begin{center}
\includegraphics[width=0.37\textwidth]{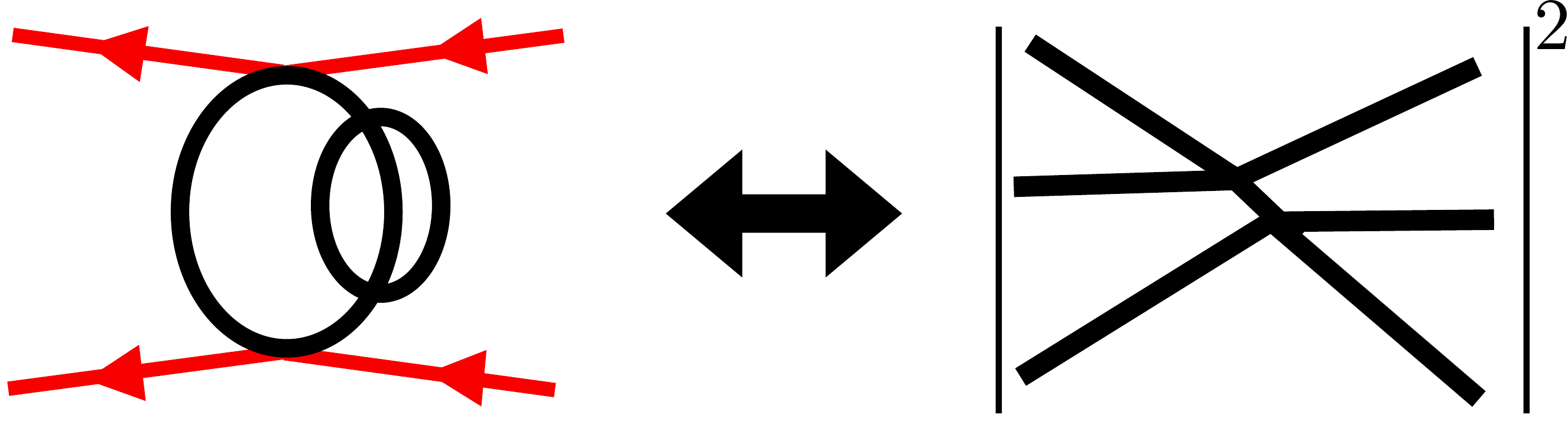}
\end{center}
\caption{Collision term at the leading order corresponding to $3\to3$ scattering.}
\label{fig:collisionTermThreeToThree}
\end{figure}

Finally, we identify diagrams contributing to the shear viscosity in the NNLO corrections.
To our knowledge, this is the first identification of the diagrams contributing to the NNLO corrections.
As noted above, the thermal mass correction to $\rho(k)$ is one of them.
The corrections in the NLO to $\mathcal{S}$ and $\mathcal{L}$
can generally contribute to the transport coefficient in the NNLO.
However, for the shear viscosity, the correction to $\mathcal{S}$ does not contribute in the NNLO. 
The reason is as follows: 
The diagrams which contribute to the vertex corrections up to of order $\lambda^2$ are shown 
in Fig.~\ref{fig:vertexCorrections}.
The second and third diagrams corresponding to the NLO corrections to $\mathcal{S}$ are
 ring diagrams, which do not depend on the external momentum $k$.
However,
the contributions of the ring diagrams vanish because
the vertex function $\mathcal{S}$ for the shear viscosity is a second rank tensor, 
and proportional to $k^ik^j-\delta^{ij}\bm{k}^2$.
Thus the corrections to $\mathcal{S}$ start from of order $\lambda^2$ that are 
higher orders of the coupling.
The diagrams of the LO and NLO corrections to ${{\fourPointVertexA}^3}_3(k,k')$ are shown in 
Fig.~\ref{fig:collisionTerm}.
These corrections correspond to quantum and thermal loop corrections of order $\lambda^3$ 
to the squared scattering amplitude of $2\to2$ process,
which do not include multiple scattering such as $3\to3$ and $2\to4$.
A typical diagram corresponding to $3\to3$ process is shown in Fig.~\ref{fig:collisionTermThreeToThree},
which process contributes to the next-to-next-to-next leading order, since the squared amplitude is proportional to $\lambda^4$.
In summary of the NNLO correction to the shear viscosity, 
the corrections of the thermal mass up to of order $\lambda^{3/2}$
and the squared scattering amplitude of order $\lambda^3$
contribute to the NNLO, while the corrections to $\mathcal{S}$ do not.

\section{Conclusion and outlook} \label{sec:Conclusion}
We have given a formulation of a resummation method for computing transport coefficients 
in relativistic quantum field theory, using the $\phi^4$ theory as an example.
We separated the diagrams into the dominant part from the other diagrams 
and reformulated them in a way which makes the included physics instructive, 
by adapting Eliashberg's method to a relativistic case in the real-time formalism.
Our result is summarized to Eqs.~(\ref{eq:vertexRen}), (\ref{eq:BetheSalpeterEquation}),
 and (\ref{eq:transportCoefficient2}).
The self-consistent equation of the vertex, Eq.~(\ref{eq:BetheSalpeterEquation}), 
has a meaning of  a kinetic equation, and has a form similar to the linearized Boltzmann equation.
In the leading order of the coupling constant, 
we recover the kinetic equation of previous works~\cite{Jeon}
(see Appendix~\ref{sec:BoltzmannVSKuboFormula}.)
The higher-order corrections beyond the leading order are systematically incorporated 
in the formalism for the first time and found to be nicely summarized
as the renormalization of the vertex correction, Eq~(\ref{eq:vertexRen}), 
the spectral function, $\rho(k)$, 
and the collision term through $\im{{\fourPointVertexA}^3}_3(k,k')$ and $\im\varPi_R(k)$. 
In the higher-orders of the collision term, 
both effects of multiple scattering and quantum loops are expressed 
 in a power of the coupling constant.
Their effects also appear in the vertex correction.
The higher-order correction is important to see the convergence of the perturbation theory 
at finite temperature.

We have identified the diagrams up to the next-to-next-leading order corrections
in the weak coupling expansion of $\phi^4$ theory.
The detailed calculation for the higher-order corrections will be discussed 
in our future work~\cite{hidaka:2011}.
We emphasize that the advantage of our diagrammatic method is to enable us 
 identify diagrams contributing to the higher-order corrections,
which is a difficult task in kinetic approaches and other diagrammatic methods.

Although the formalism is developed 
 using the $\phi^4$ theory in the present work,
the generalization of it to fermionic theories is straight forward:
The Bose-Einstein distribution function in the self-consistent equation 
is simply replaced by the Fermi-Dirac distribution function.
The spinor structure is introduced into the vertex function and the collision term.
Since the Fermi-Dirac distribution function $n_F(E)$ does not diverge at $E=0$, 
a infrared cutoff introduced in Sec.~\ref{oneLoopAnalysis} is not necessary.

Gauge theory is more complicated. Resummation of collinear divergences
in addition to the ladder diagrams is necessary to obtain the correct result 
in certain orders of the coupling constant,
which is called Landau-Pomeranchuk-Migdal (LPM) effect~\cite{LPM}. 
The self-consistent equation summing the collinear singularities over
has the form of the Boltzmann equation with the collision term in $1\to2$ and $2\to1$ process~\cite{amy,Gagnon,Arnold:2002zm}.

It is also interesting to apply the diagram method to critical phenomena,
where the hydrodynamic mode and the fluctuation of order parameter play an important role. 
We have to take into account these modes in addition to quasiparticle modes.

\acknowledgments
This work was partially supported by a
Grant-in-Aid for Scientific Research by the Ministry of Education,
Culture, Sports, Science and Technology (MEXT) of Japan (No.
20540265 and No. 19$\cdot$07797),
 by Yukawa International Program for Quark-Hadron Sciences, and by the
Grant-in-Aid for the global COE program ``The Next Generation of
Physics, Spun from Universality and Emergence'' from MEXT.

\appendix

\section{Several Identities in the real-time formalism}
\label{sec:identities}
In this Appendix, we show several useful identities in the $R/A$ basis \cite{vanEijck:1994rw}.
We start with the standard basis.
The largest and smallest time equation shows
\begin{equation}
\sum_{\{i_{k}\}} \varGamma_{i_{1}i_{2}\cdots i_{n}}(p_{1},p_{2},\cdots,p_{n}) \prod_{i_{k}=2} e^{-\sigma p^{0}_{k}}=0 \,,
\label{eq:largestTime}
\end{equation}

and Kubo-Martin-Schwinger (KMS) relation shows
\begin{equation}
\sum_{\{i_{k}\}} \varGamma_{i_{1}i_{2}\cdots i_{n}}(p_{1},p_{2},\cdots,p_{n}) \prod_{i_{k}=2} e^{(\beta-\sigma )p^{0}_{k}}=0 \,.
\label{eq:KMS}
\end{equation}
By diagrammatic analysis, one finds \cite{vanEijck:1994rw}
\begin{equation}
\begin{split}
&\varGamma^{*}_{i_{1}i_{2}\cdots i_{n}}(p_{1},p_{2},\cdots,p_{n}) \\
&\quad=
-\varGamma_{\bar{i}_{1}\bar{i}_{2}\cdots \bar{i}_{n}}(p_{1},p_{2},\cdots,p_{n}) \prod_{i_{k}=2} e^{-(\beta-2\sigma )p^{0}_{k}} \,.
\end{split}
\label{eq:complexConjugate}
\end{equation}
Equations~(\ref{eq:largestTime}) to (\ref{eq:complexConjugate}) are basic identities in the real-time formalism~\cite{Kobes:1985kc}.
In the $R/A$ basis, Eq.~(\ref{eq:largestTime}) becomes
\begin{equation}
\begin{split}
&\sum_{\{i_{k}\}}  \prod_{m=1}^n{{U}^{\alpha_{m}}}_{i_{m}}(p_{m}) \prod_{i_{k}=2} e^{-\sigma p^{0}_{k}}
\varGamma_{\alpha_{1}\alpha_{2}\cdots \alpha_{n}}(p_{1},p_{2},\cdots,p_{n}) \\
&\qquad=  \prod_{m=1}^n\Big[ {{U}^{\alpha_{m}}}_{1}(p_{m})+{{U}^{\alpha_{m}}}_{2}(p_{m})e^{-\sigma p^{0}_{m}} \Bigr]\\
&\qquad\qquad\times\varGamma_{\alpha_{1}\alpha_{2}\cdots \alpha_{n}}(p_{1},p_{2},\cdots,p_{n}) \\
&\qquad=
\varGamma_{RR\cdots R}(p_{1},p_{2},\cdots,p_{n})\prod_{m=1}^n( e^{\beta p^0_m}-1 )=0  \,.
\end{split}
\end{equation}
Since $e^{\beta p^0_m}-1\neq0$, 
\begin{equation}
\varGamma_{RR\cdots R}(p_{1},p_{2},\cdots,p_{n})=0 \,.
\end{equation}
Similarly, Eq.~(\ref{eq:KMS}) becomes in this basis
\begin{equation}
\varGamma_{AA\cdots A}(p_{1},p_{2},\cdots,p_{n})=0  \,.
\end{equation}
Therefore, the vertex function vanishes, when all indices are the same.
This means particles are not produced from the thermal equilibrium system nor absorbed into the system,
similar to the vacuum at zero temperature and zero density.

Equation~(\ref{eq:complexConjugate}) is reduced to 
\begin{equation}
\begin{split}
&\varGamma^{*}_{\alpha_{1}\alpha_{2}\cdots \alpha_{n}}(p_{1},p_{2},\cdots,p_{n})\\
&\quad=\varGamma^{*}_{i_{1}i_{2}\cdots i_{n}}(p_{1},p_{2},\cdots,p_{n}) \prod_{m=1}^n{U^{i_{m}}}_{\alpha_{m}}(p_{m})\\
&\quad=-\varGamma_{\bar{\alpha}_{1}\bar{\alpha}_{2}\cdots \bar{\alpha}_{n}}(p_{1},p_{2},\cdots,p_{n})\\
&\quad\qquad\times\prod_{\alpha_{m}=R} f(p_{m}^{0})\prod_{\alpha_{m}=A}\frac{1}{-f(-p_{m}^{0})} \,.
\end{split}
\end{equation}
For a two-point function,
\begin{equation}
\varGamma^{*}_{AR}(-k,k)=\varGamma_{RA}(-k,k) \,,
\end{equation}
implying $\DR(k)={\DA}^*(k)$.
For a three-point function,
\begin{equation}
\varGamma^{*}_{RAA}(p,q,k)
=-\varGamma_{ARR}(p,q,k)\frac{1}{f(p+k)-f(k)} \,.
\end{equation}
Therefore,
\begin{equation}
-({f(p+k)-f(k)})\varGamma^{*}_{RAA}(p,q,k)=\varGamma_{ARR}(p,q,k) \,.
\end{equation}
For a four-point function,
\begin{equation}
\begin{split}
&{\fourPointVertex}^{*}_{AARR}(p_{1},p_{2},p_{3},p_{4})\\
&\quad= -{\fourPointVertex}_{RRAA}(p_{1},p_{2},p_{3},p_{4})
\frac{f(p^{0}_{3})f(p^{0}_{4})}{f(-p^{0}_{1})f(-p^{0}_{2})} \,.
\label{eq:IdentityFourPointFunction}
\end{split}
\end{equation}

\section{Relation between Boltzmann equation and Kubo formula in the leading order}
\label{sec:BoltzmannVSKuboFormula}
Here we explicitly show that Eq.~(\ref{eq:collisionOperator}) is equivalent to a linearized Boltzmann equation in the leading order. 
This was shown in Ref.~\cite{Jeon} in a different formalism.
Let us start with a Boltzmann equation,
\begin{equation}
2k^{\mu}\partial_{\mu} f(x,k)
=-C[f]\,.	
\label{eq:BoltzmannEquationClassical}
\end{equation}
where $f(x,k)$ denotes the distribution function, and  $C[f]$ is the collision term. In the leading order,  the collision term has the form of 
$2\to2$ collision,
\begin{equation}
\begin{split}
C[f]&=\frac{1}{2}\int\frac{d^3k'}{(2\pi)^3}\frac{1}{2E_{k'}}
\int\frac{d^3q}{(2\pi)^3}\frac{1}{2E_{q}}
\int\frac{d^3q'}{(2\pi)^3}\frac{1}{2E_{q'}}\\
&\quad\times |\mathcal{M}|^2(2\pi)^4\delta^{(4)}(k+q-k'-q')\\
&\qquad\times\bigl[f_k f_q(1+f_{k'})(1+f_{q'})\\
&\qquad\qquad- (1+f_k)(1+f_{q})f_{k'}f_{q'}\bigr] \,,
\end{split}
\end{equation}
where $\mathcal{M}$ is the scattering amplitude; $\mathcal{M}=i\lambda$ for $\phi^4$ theory in the leading order.
At (local) thermal equilibrium, the distribution function has the form
\begin{equation}
f_0(x,k) = \frac{1}{\exp\bigl({\beta(x) u_\mu(x) k^\mu}\bigr)-1} \,,
\end{equation}
where $u_\mu(x)$ is the local velocity, and $k^\mu$ is the four momentum at on-shell, $k^2=m^2$, with mass $m$.
Here we consider the shear flow, then the LHS of Eq.~(\ref{eq:BoltzmannEquationClassical}) becomes
\begin{equation}
2k^\mu \partial_\mu f_0(x,p)= -
\beta f_0(x,p)\bigl(1+f_0(x,p)\bigr)\varGamma^{ij}(k)\frac{1}{2}\sigma_{ij}(x) \,,
\end{equation}
where $\varGamma^{ij}(k)=2k^i k^j-2\delta^{ij}\bm{k}^2/3$, and
\begin{equation}
\sigma_{ij} = \partial_i u_j(x)+\partial_j u_i(x)-\frac{2}{3}\delta_{ij}\partial_k u_k \,.
\end{equation}
 We have considered the local rest frame, $u_\mu = (1,\bm{0})$. Note that at the local rest frame $\partial_i u_j(x)$ is nonzero,
although $u_i(x)=0$.
By linearizing the Boltzmann equation around the thermal equilibrium $f(x,k)=f_0(x,k)+\delta f(x,k)$, one finds
\begin{equation}
\begin{split}
C[f]&\simeq\beta\bigl(1+f_0(E_k)\bigr)\frac{\lambda^2}{2}\int\frac{d^3k_q}{(2\pi)^3}\frac{1}{2E_q}\bigl(1+f_0(E_q)\bigr)\\
&\quad\times\int\frac{d^3{k'}}{(2\pi)^3}\frac{1}{2E_{k'}}f_0(E_{k'})
\int\frac{d^3k_{q'}}{(2\pi)^3}\frac{1}{2E_{q'}}f_0(E_{q'}) \\
&\qquad\times (2\pi)^4\delta^{(4)}(k+q-k'-q')\\
&\quad\qquad\times\bigl[\varphi^{ij}(k)+\varphi^{ij}(q)-\varphi^{ij}(k')-\varphi^{ij}(q')\bigr]\frac{1}{2}\sigma_{ij}\\
&\equiv2E_k\beta f_0(E_k)\bigl(1+f_0(E_k)\bigr)\mathcal{L}_\text{Boltz}\varphi^{ij}\frac{1}{2}\sigma_{ij}\,,
\end{split}
\label{eq:collisonTermBoltzman}
\end{equation}
where $f_0(E_q)\equiv f_0(x,k)$ at the local rest frame, $\mathcal{L}_\text{Boltz}$ is  the linearized collision operator,
and we chose $\delta f(x,k)=\beta f_0(E_k)(1+f_0(E_k))\varphi^{ij}(k)\sigma_{ij}/2$.
Then, the linearized equation reads
\begin{equation}
\frac{\varGamma^{ij}(k)}{2E_k} = \mathcal{L}_\text{Boltz}\varphi^{ij}(k) \,.
\label{eq:liniearizedBoltzmannClassical}
\end{equation}
This is the same form as Eq.~(\ref{eq:linearizedBoltzmannEquation}).
Let us check $\mathcal{L} = \mathcal{L}_\text{Boltz.}$ at weak coupling. For this purpose, let us estimate the four-point function, $\im{{\fourPointVertexA}^{3}}_{3}(k,k')$, and $\im \varPi^\text{two-loop}_{R}(k)$.
Using Feynman rule, Eqs.~(\ref{eq:feynmanRuleDR}) to (\ref{eq:feynmanRuleFourPoint}), we find
\begin{equation}
\begin{split}
&\im{{\fourPointVertexA}^{3}}_{3}(k,k')\\
&\quad=-\frac{1}{2}\lambda^{2}\bigl(e^{\beta k^{0}}-1\bigr)
f(k'^{0})
\int\frac{d^{4} q}{(2\pi)^{4}}\bigl(1+f(q^{0})\bigr)\rho(q) \\
&\qquad\times\int\frac{d^{4} q'}{(2\pi)^{4}}f(q'^{0})\rho(q') 
(2\pi)^{(4)}\delta(k+q-k'-q') \,,
\end{split}
\label{eq:T33B}
\end{equation}
and
\begin{equation}
\begin{split}
&\im \varPi^\text{two-loop}_{R}(k)\\
&\quad=-\bigl(e^{\beta k^{0}}-1\bigr)\frac{\lambda^{2}}{12}
\int\frac{d^{4}q}{(2\pi)^{4}}\bigl(1+f(q^0)\bigr)\rho(q)\\
&\qquad\times\int\frac{d^{4}q'}{(2\pi)^{4}}f(q'^{0})\rho(q')
\int\frac{d^{4}k'}{(2\pi)^{4}}f(k'^{0})\rho(k')\\
&\qquad\qquad\times(2\pi)^{4}\delta^{(4)}(k+q-k'-q') \,.
\end{split}
\label{eq:ImSelfEnergy}
\end{equation}
Comparing Eqs.~(\ref{eq:T33B}) and (\ref{eq:ImSelfEnergy}), we find
\begin{equation}
\im \varPi^\text{two-loop}_{R}(k)
=\frac{1}{6}\int\frac{d^{4}k'}{(2\pi)^{4}}\rho(k') \im {{\fourPointVertexA}^{3}}_{3}(k,k') \,.
\end{equation}
The collision term becomes
\begin{equation}
\begin{split}
\mathcal{L}\varphi^{ij}(k)&=
-2\,\frac{1}{2k^0}\,\im\varPi^\text{two-loop}_{R}(k)\varphi^{ij}(k)\\
&\quad+\frac{1}{2k^0}\int \frac{d^{4}k'}{(2\pi)^{4}}\rho(k')
\im{{\fourPointVertexA}^{3}}_{3}(k,k')
\varphi^{ij}(k') 
\\
&= \,\frac{f^{-1}(k^{0})}{2k^0}\frac{\lambda^{2}}{6}
\int\frac{d^{4}k'}{(2\pi)^{4}}f(k'^{0})\rho(k')\\
&\quad\times\int\frac{d^{4}q}{(2\pi)^{4}}\bigl(1+f(q^{0})\bigr)\rho(q)
\int\frac{d^{4}q'}{(2\pi)^{4}}f(q'^{0})\rho(q') \\
&\qquad\times(2\pi)^4\delta^{(4)}(k+q-k'-q')\\
&\quad\qquad\times\bigl[\varphi^{ij}(k)-\varphi^{ij}(k')+\varphi^{ij}(q)-\varphi^{ij}(q')\bigr] \,.
\end{split}
\label{eq:collisionTemLeading}
\end{equation}
Here we have used the relation $\varphi^{ij}(-k)=-\varphi^{ij}(k)$ derived from Eq.~(\ref{eq:phi}) to obtain the third line.
Equation~(\ref{eq:collisionTemLeading}) includes positive and negative energies, while 
Eq.~(\ref{eq:collisonTermBoltzman}) includes the only positive energy.
This collision term contains three processes: $1\to3$, $2\to2$, and $3\to1$ scatterings.
$1\to3$ and $3\to1$ collisions can be neglected in the leading order 
because the on-shell condition of quasiparticle does not satisfy.
Then, the collision term for the positive energy state, $k^0>0$, becomes
\begin{equation}
\begin{split}
\mathcal{L}\varphi^{ij}(k)
&=\frac{f^{-1}(k^{0})}{2k^0}\frac{\lambda^{2}}{2}
\int\frac{d^{4}k'}{(2\pi)^{4}}f(k'^{0})\rho(k')\theta(k'^0)\\
&\quad\times\int\frac{d^{4}q}{(2\pi)^{4}}\bigl(1+f(q^{0})\bigr)\rho(q)\theta(q^0)\\
&\qquad\times\int\frac{d^{4}q'}{(2\pi)^{4}}f(q'^{0})\rho(q')\theta(q'^0) \\
&\quad\qquad\times(2\pi)^4\delta^{(4)}(k+q-k'-q')\\
&\qquad\qquad\times\bigl[\varphi^{ij}(k)-\varphi^{ij}(k')+\varphi^{ij}(q)-\varphi^{ij}(q')\bigr] 
\,.
\end{split}
\end{equation}
By using the quasiparticle approximation, 
\begin{equation}
\rho(k)\theta(k^0)=\frac{1}{2E_k}(2\pi)\delta(k^0-E_k),
\end{equation}
with $E_k=\sqrt{\bm{k}^2+m_T^2}$,
we obtain the collision operator as
\begin{equation}
\begin{split}
&\mathcal{L}\varphi^{ij}(k)\\
&\quad= \frac{f^{-1}(E_k)}{2E_k}\frac{\lambda^{2}}{2}
\int\frac{d^{3}k'}{(2\pi)^{3}}\frac{1}{2E_{k'}}f(E_{k'})\\
&\quad\qquad\times\int\frac{d^{3}q}{(2\pi)^{3}}\frac{1}{2E_{q}}\bigl(1+f(E_{q})\bigr)
\int\frac{d^{3}q'}{(2\pi)^{3}}\frac{1}{2E_{q'}}f(E_{q'})\\
&\quad\qquad\qquad\times(2\pi)^4\delta^{(4)}(k+q-k'-q')\\
&\qquad\qquad\qquad\times\bigl[\varphi^{ij}(k)-\varphi^{ij}(k')+\varphi^{ij}(q)-\varphi^{ij}(q')\bigr] 
\,.
\end{split}
\end{equation}
Thus, the collision operator in the Boltzmann equation is equivalent to that in diagrammatic method in the leading order,
$\mathcal{L}=\mathcal{L}_\text{Boltz.}$. 
In the neutral scalar theory, the negative energy state is identical to the positive energy state, so that the collision term for the negative energy state is the same as that for the positive energy state. 

\section{Some properties of the collision operator}
The collision operator of the linearized Boltzmann equation has some basic properties: semi-positive definiteness, self-adjointness, and the conserved charges being the collision invariants.
Here, we focus on the self-adjointness and the conserved charges being the collision invariants,
although the semi-positive definiteness of the collision operator is necessary to ensure the stability of the thermal equilibrium. 
\subsection{Self-adjointness of the collision operator}
In order to show the self-adjointness, we write the linearized collision operator as
\begin{equation}
\mathcal{L}\varphi(k)=\int\frac{d^{4}k'}{(2\pi)^{4}}W(k') \mathcal{L}(k,k')\varphi(k') \,,
\end{equation}
where
\begin{equation}
\begin{split}
\mathcal{L}(k,k')& = 
\frac{-1}{k^0}\,\im\varPi_{R}(k)\,W^{-1}(k')\delta(k-k')\\
&\qquad+  \frac{1}{2k^0}\,\im{{\fourPointVertexA}^{3}}_{3}(k,k')
W^{-1}(k')\rho(k')\,.
\end{split}
\label{eq:collsionOperator}
\end{equation}
This is the momentum representation of the linearized collision operator.
The Hermite conjugate is defined as
\begin{equation}
\langle \varphi_{1},\mathcal{L}\varphi_{2}\rangle=\langle\mathcal{L}^{\dagger}\varphi_{1},\varphi_{2}\rangle \,,
\end{equation}
which means that
\begin{align}
\begin{split}
\langle\varphi_{1},\mathcal{L}\varphi_{2}\rangle&=
\int\frac{d^{4}k}{(2\pi)^{4}}W(k)\varphi^{*}_{1}(k)\\
&\quad\times\left[\int\frac{d^{4}k'}{(2\pi)^{4}}W(k') \mathcal{L}(k,k')\varphi_{2}(k') \right]\\
&=
\int\frac{d^{4}k'}{(2\pi)^{4}}W(k')\\
&\quad\times\left[\int\frac{d^{4}k}{(2\pi)^{4}}W(k)
\mathcal{L}^{*}(k,k')\varphi_{1}(k)\right]^{*} \varphi_{2}(k')\,,
\end{split}\\
\begin{split}
\langle \mathcal{L}^\dag\varphi_1,\varphi_2\rangle&=
\int\frac{d^{4}k'}{(2\pi)^{4}}W(k')\\
&\quad\times\left[\int\frac{d^{4}k}{(2\pi)^{4}}W(k)
 \mathcal{L}^{\dagger}(k',k)\varphi_{1}(k)\right]^{*} \varphi_{2}(k') \,,
 \end{split}
\end{align}
or
\begin{equation}
\mathcal{L}^{\dagger}(k',k)= \mathcal{L}^{*}(k,k') \,.
\end{equation}
Let us check the self-adjointness or Hermiticity, $\mathcal{L}^\dagger=\mathcal{L}$. 
The first term of RHS in Eq.~(\ref{eq:collsionOperator}) is obviously symmetric under $k\leftrightarrow k'$. 
The four-point function, ${{\fourPointVertexA}^{3}}_{3}(k,k')$, in Eq.~(\ref{eq:collsionOperator}) has a symmetry,
\begin{equation}
{{\fourPointVertexA}^{*3}}_{3}(k,k')
=-{{\fourPointVertexA}^{3}}_{3}(k',k)\frac{f(k')\bigl(1+f(k')\bigr)}{f(k)\bigl(1+f(k)\bigr)} \,,
\end{equation}
from Eq.~(\ref{eq:IdentityFourPointFunction}).
Thus the second term of RHS in Eq.~(\ref{eq:collsionOperator}) is symmetric:
\begin{equation}
\begin{split}
&\im{{\fourPointVertexA}^{3}}_{3}(k,k')f^{-1}(k'^0)\bigl(1+f(k'^0)\bigr)^{-1}\\
&\qquad=\,\im{{\fourPointVertexA}^{3}}_{3}(k',k)f^{-1}(k^0)\bigl(1+f(k^0)\bigr)^{-1}\,.
\end{split}
\end{equation}
Therefore, the collision operator is Hermite, $\mathcal{L}^{\dagger}=\mathcal{L}$.
More precisely, the collision operator is symmetric because Eq.~(\ref{eq:collsionOperator}) is a real function.

\subsection{Collision invariant and Ward-Takahashi identity}\label{sec:WTIdentity}
In this subsection, we derive Ward-Takahashi (WT) identity in the $R/A$ basis, 
and show the WT identity implies that the conserved charges are zero modes of the collision operator, Eq.~(\ref{eq:collisionOperator}).
In the imaginary-time formalism, WT identity is derived in Refs.~\cite{Toyoda:1986jn,Aarts:2002tn,Gagnon}.
The WT identity is useful to constrain diagrams in order to keep symmetries in the resummed perturbation theory~\cite{Gagnon}.
Conversely, to obtain the correct result in a certain order of perturbation theory, the collision term must satisfy the WT identity.

The conserved current in $\phi^4$ theory is a energy-momentum current, defined by 
\begin{equation}
j_{\nu}^{\mu}=T_{\nu}^{\mu}=\partial_{\nu} \phi \partial^{\mu} \phi-\delta^{\mu}_{\nu}\mathcal{L}\,.
\end{equation}
In particular, the charge density of the momentum current is the bilinear operator,
\begin{equation}
j_{i}^{0} = \partial_{i}\phi\partial^{0}\phi \,.
\end{equation}
This bilinear form is important to relate the WT identity and collision term, as we will see later.
In order to derive the WT identity in thermal bath, 
consider the following correlation function:
\begin{equation}
\langle T_{C} j_{i}^{\mu}(z)\phi(x)\phi(y) \rangle \,,
\end{equation}
where $T_C$ denotes a complex-time-path ordering shown in Fig.~\ref{fig:timepath}.
Taking $\partial_{z\mu}$, we obtain
\begin{equation}
\begin{split}
&\partial_{z\mu}\langle T_{C} j_{i}^{\mu}(z)\phi(x)\phi(y) \rangle\\
&\quad= \delta_{C}(z^{0}-x^{0}) \langle T_{C}[j_{i}^{0}(z),\phi(x)]\phi(y)\rangle \\
&\qquad+\delta_{C}(z^{0}-y^{0})\langle T_{C}\phi(x) [j_{i}^{0}(z),\phi(y)]\rangle  \,,
\end{split}
\end{equation}
where we used the conservation law $\partial_\mu j^\mu_i=0$.
Noting that the equal-time commutation relation gives an infinitesimal translation,
\begin{equation}
[j_{i}^{0}(z),\phi(x)]=-i\delta^{(3)}(\bm{z}-\bm{x})\partial_{i}\phi(x) \,,
\end{equation}
we obtain
\begin{equation}
\begin{split}
\partial_{z\mu}\langle T_{C} j_{i}^{\mu}(z)\phi(x)\phi(y) \rangle
&=-i \delta^{(4)}_{C}(z-x) \langle T_{C} \partial_{i}\phi(x)\phi(y)\rangle \\
&\quad-i\delta_{C}^{(4)}(z-y) \langle T_{C}\phi(x) \partial_{i}\phi(y)\rangle  \,.
\end{split}
\label{eq:WTIdentity1}
\end{equation}
This is the WT identity in the coordinate space on the complex time path.
In the real-time formalism, the complex path is decomposed into
 four parts as shown in Fig.~\ref{fig:timepath}.
In the standard basis, Eq.~(\ref{eq:WTIdentity1}) becomes
\begin{equation}
\begin{split}
&\partial_{z\mu}\langle T_{C} j_{i;c}^{\mu}(z)\phi^{a}(x)\phi^{b}(y) \rangle\\
&\quad=-i{X_{cd}}^{a} \delta^{(4)}(z-x) \langle T \partial_{zi}\phi^{d}(z)\phi^{b}(y)\rangle  \\
&\quad\qquad-i{{X_c}^b}_d\delta^{(4)}(z-y) \langle T\phi^{a}(x) \partial_{zi}\phi^{d}(z)\rangle  \,,
\end{split}
\label{eq:WTIdentity2}
\end{equation}
where $j_{i;c}^{0}(x)=X_{cde}\partial_{i}\phi^{d}(x)\partial^{0}\phi^{e}(x)$ with $X_{cde}=\delta_{cd}g_{de}$.
Using the propagator and  the three-point vertex $\varGamma^\mu(z,w,w')$, we obtain
\begin{equation}
\begin{split}
&\partial_{z\mu} \int d^{4}w\int d^{4}w' \varGamma^{\mu}_{i;cde}(z,w,w')
D^{da}(w,x)D^{eb}(w',y)\\
&\quad=-i{X_{cd}}^{a}  \delta^{(4)}(z-x)\partial_{zi}D^{db}(z,y) \\
&\quad\qquad-i{{X_c}^b}_d \delta^{(4)}(z-y)\partial_{zi}D^{da}(z,x).
\end{split}
\end{equation}
In momentum space, this becomes
\begin{equation}
\begin{split}
&ip_{\mu}\varGamma^{\mu}_{i;cde}(p,q,k)D^{da}(q)D^{eb}(k)\\
&\qquad=
-  k_{i}{X_{cd}}^{a} D^{db}(k)-  q_{i} {{X_c}^{b}}_dD^{da}(q) \,,
\end{split}
\label{eq:WTmometum}
\end{equation}
where $p$, $k$ and $q$ are not independent but satisfies $p+k+q=0$.
Multiplying both sides of Eq.~(\ref{eq:WTmometum}) by $D^{-1}_{aa'}(q)D^{-1}_{bb'}(k)$, we find

\begin{equation}
ip_{\mu}\varGamma^{\mu}_{i;ca'b'}(p,q,k)=
- k_{i}{X_{cb'}}^{d} D^{-1}_{da'}(q) 
-q_{i} {{X_c}^{d}}_{a'} D^{-1}_{db'}(k)  \,.
\end{equation}
At tree level, this is reduced to
\begin{equation}
ip_{\mu}\varGamma^{\mu}_{(0)i;cab}(p,q,k)=
- k_{i} {X_{cb}}^{d} D^{-1}_{F;da}(q) 
-q_{i} {{X_c}^{d}}_a D^{-1}_{F;cb}(k)   \,,
\end{equation}
where $\varGamma^{\mu}_{(0)i;cab}(p,q,k)$ is the tree-level vertex,
\begin{equation}
\varGamma^{\mu}_{(0)i;cab}(p,q,k)=X_{cab}\bigl(-(q_ik^{\mu}+k_iq^{\mu} )+\delta^{\mu}_{i}(k\cdot q+m^{2}) \bigr)\,,
\end{equation}
and $D_{F}^{ab}(k)$ is the free propagator.
Here, we decompose the full vertex-function into that at tree level and its correction,
$\delta{\varGamma}_{i;abc}^{\mu}$:
\begin{equation}
\varGamma_{i;abc}^{\mu}(p,q,k)\equiv \varGamma^{\mu}_{(0)i;abc}(p,q,k)
+ \delta\varGamma^{\mu}_{i;abc}(p,q,k) \,.
\end{equation}
Then, the correction to the vertex satisfies 
\begin{equation}
ip_{\mu}\delta\varGamma^{\mu}_{i;cab}(p,q,k)=
-ik_{i}{X_{cb}}^{d}\varPi_{da}(q) 
-iq_{i}{{X_c}^{d}}_a\varPi_{db}(k) \,.
\label{eq:WTCorrection}
\end{equation}
In the vacuum, Eq~(\ref{eq:WTCorrection}) gives the relation between the charge and 
 wave function renormalization factors. 
 In addition to that, at finite temperature, we will show that conserved charges are the collision invariants
 of the collision operator. To see this, we, first, take the limit $\bm{p}\to\bm{0}$,  
\begin{equation}
\begin{split}
ip_{0}\delta{\varGamma}^{0}_{i;cab}(p,q,k)&=
-ik_{i} \bigl({X_{cb}}^{d}\varPi_{da}(q) 
-{{X_c}^{d}}_a\varPi_{db}(k) \bigr)\\
& \equiv F_{i;cab}(k,p^0) \,.
 \end{split}
 \label{eq:Fcab1}
\end{equation}
As we mentioned above, $j^{0}_{i}$ is a bilinear operator, so that the three-point vertex can be written by using the four-point vertex ${\fourPointVertex}_{abcd}$ as 
\begin{equation}
\begin{split}
\delta{\varGamma}^{0}_{i;cab}(p,q,k)&= \int\frac{d^{4}k'}{(2\pi)^{4}}
(-i){\fourPointVertex}_{abde}(q,k,-k',-q')\\
&\qquad\times D^{fd}(k')D^{ge}(q')\varGamma_{(0)i;cgf }^{0}(p,q',k') \,,
\end{split}
\end{equation}
so that  $F_{i;cab}(k,p^0)$ becomes
\begin{equation}
\begin{split}
F_{i;cab}(k,p^0)
&=
 -i\int\frac{d^{4}k'}{(2\pi)^{4}}{\fourPointVertex}_{abde}(q,k,-k',-q')\\
&\qquad\times
k'_{i}\bigl({{X_c}^{d}}_gD^{ge}(q')- {X_{cf}}^{e}D^{fd}(k')\\
&\qquad\quad-D^{fd}(k')D^{ge}(k')F_{i;cgf}(k',p^0) \bigr) \,.
\end{split}
\label{eq:Fcab2}
\end{equation}
Let us define ${{\fourPointVertexA}}_{abde}(q,k,-k',-q')$ by
\begin{equation}
\begin{split}
&{\fourPointVertex}_{abde}(q,k,-k',-q')\\
&\quad={{\fourPointVertexA}}_{abde}(q,k,-k',-q')\\
&\qquad+\int\frac{d^{4}k''}{(2\pi)^{4}}(-i){{\fourPointVertex}}_{abd'e'}(q,k,-k'',-q'') \\
&\quad\qquad\times D^{b'd'}(k'')D^{a'e'}(q''){{\fourPointVertexA}}_{a'b'de}(q'',k'',-k',-q') \,.
\end{split}
\label{eq:Ttilde}
\end{equation}
This is equivalent to Eq.~(\ref{eq:selfConsistentEqForFourPointVertex}) but in the standard basis.
Comparing Eq.~(\ref{eq:Ttilde}) with Eq.~(\ref{eq:Fcab2}), we find
\begin{equation}
\begin{split}
F_{i;cab}(k,p^0)=&
-i \int\frac{d^{4}k'}{(2\pi)^{4}}k'_{i} {{\fourPointVertexA}}_{abde}(q,k,-k',-q')\\
&\qquad\times\bigl({{X_c}^{d}}_gD^{ge}(q') - {X_{cf}}^{e}D^{fd}(k')\bigr) \,.
\end{split}
\label{eq:Fcab3}
\end{equation}
We have derived Eq.~(\ref{eq:Fcab3}) in the standard basis; however, this is independent of a choice of basis. 
Now, let us apply it to the $R/A$ basis. We take the case $a=b=A$, and $c=R$:
\begin{align}
{X_{RA}}^{R}(p,q,k)&=
1 \,,\\
{X_{RA}}^{A}(p,q,k)&=
1+f(p^{0})+f(k^{0})\,,\\
{X_{RR}}^{R}(p,q,k)&=
1+f(p^{0})+f(q^{0}) \,,\\
{X_{RR}}^{A}(p,q,k)&=
0 \,.
\end{align}
Therefore, $F_{i;RAA}$ becomes
\begin{equation}
\begin{split}
&F_{i;RAA}(k,p^0)\\
&\quad=
-i \int\frac{d^{4}k'}{(2\pi)^{4}}k'_{i}\Bigl[
{{\fourPointVertexA}}_{AARR}(-p-k,k,-k',p+k')\\
&\quad\qquad\times(-i)\bigl(\DR(p+k') - \DA(k')\bigr) \\
&\qquad+{{\fourPointVertexA}}_{AARA}(-p-k,k,-k',p+k')
f(k^0)\\
&\quad\qquad\times(-i)\bigl( \DA(p+k') - \DA(k')\bigr) \\
&\qquad-{{\fourPointVertexA}}_{AAAR}(-p-k,k,-k',p+k')
f(k^0+p^0)\\
&\qquad\quad\times(-i)\bigl( \DR(p+k') - \DR(k')\bigr) 
\Bigr] \,,
\end{split}
\end{equation}
where we have omitted $1+f(p^0)$  in the second (third) term 
because there is no pole in the lower (upper) complex $k'^0$ plane.
Taking the limit $p^0\to0$, we find
\begin{equation}
\begin{split}
&\lim_{p_0\to0}F_{i;RAA}(k,p^0)\\
&\quad=
-i \int\frac{d^{4}k'}{(2\pi)^{4}}k'_{i}
{{\fourPointVertexA}}_{AARR}(-k,k,-k',k')
\rho(k') \,.
\end{split}
\label{eq:Fraa0}
\end{equation}
On the other side, from Eq.~(\ref{eq:Fcab1}),
\begin{equation}
\begin{split}
F_{i;RAA}(k,p^0)&=-ik_{i}\bigl({X_{RA}}^{R}(p,q,k)\varPi_{RA}(q) \\
&\quad\qquad-{{X_{R}}^{R}}_{A}(p,q,k)\varPi_{RA}(k)\bigr)\\
&=-ik_{i}\bigl(\varPi_{A}(q) -\varPi_{A}(k) \bigr) \,.
\end{split}
\end{equation}
Taking $p^0\to0$, we find
\begin{equation}
\lim_{p^0\to0}F_{i;RAA}(k,p^0)=2\,\im\varPi_{R}(k)\, k_{i} \,.
\label{eq:Fraa1}
\end{equation}
Inserting Eq.~(\ref{eq:Fraa1}) into Eq.~(\ref{eq:Fraa0}),
we find
\begin{equation}
 - 2\,\im\varPi_{R}(k)\, k_{i}
+ \int\frac{d^{4}k'}{(2\pi)^{4}}\im{{{\fourPointVertexA}}^3}_{3}(k,k') \rho(k')
k'_i =0 \,,
\end{equation}
where $\im{{{\fourPointVertexA}}^3}_{3}(k,k') = \im{{\fourPointVertexA}}_{AARR}(-k,k,-k',k')$ is used.
Using the collision operator, Eq.~(\ref{eq:collisionOperator}), we find
\begin{equation}
\mathcal{L}k_{i} =0 \,.
\end{equation}
Therefore, the conserved charge $k_i$ is zero mode of the collision operator.


\begin{thebibliography}{99}

\bibitem{whitepaper}
J.~Adams {\em et~al.}, 
\mydoi{10.1016/j.nuclphysa.2005.03.085}{\npa{757}{102}{2005}}
\arXiv{nucl-ex/0501009}{[arXiv:nucl-ex/0501009]};
%
K.~Adcox {\em et~al.}, 
\mydoi{10.1016/j.nuclphysa.2005.03.086}{\ibid{757}{184}{2005}}
\arXiv{nucl-ex/0410003}{[arXiv:nucl-ex/0410003]};
%
I.~Arsene {\em et~al.}, 
\mydoi{10.1016/j.nuclphysa.2005.02.130}{\ibid{757}{1}{2005}}
\arXiv{nucl-ex/0410020}{[arXiv:nucl-ex/0410020]};
%
B.~B. Back {\em et~al.}, 
\mydoi{10.1016/j.nuclphysa.2005.03.084 ]}{\ibid{757}{28}{2005}}
\arXiv{nucl-ex/0410022}{[arXiv:nucl-ex/0410022]}.
%
%
\bibitem{strong}
M.~Gyulassy and L.~McLerran,
\mydoi{10.1016/j.nuclphysa.2004.10.034}{\npa{750}{30}{2005}}
\arXiv{nucl-th/0405013}{[arXiv:nucl-th/0405013]};
%
A.~Peshier and W.~Cassing,
\mydoi{10.1103/PhysRevLett.94.172301}{\prl{94}{172301}{2005}}
\arXiv{hep-ph/0502138}{[arXiv:hep-ph/0502138]};
%
B.~Muller and J.~L.~Nagle,
\mydoi{10.1146/annurev.nucl.56.080805.140556}{\arnps{56}{93}{2006}}
\arXiv{nucl-th/0602029}{[arXiv:nucl-th/0602029]};
%
S.~Mrowczynski and M.~H.~Thoma,
\mydoi{10.1146/annurev.nucl.57.090506.123124}{\ibid{57}{61}{2007}}
\arXiv{nucl-th/0701002}{[arXiv:nucl-th/0701002]};
%
E. V. Shuryak,
\mydoi{10.1016/j.ppnp.2008.09.001}{Prog.\ Part.\ Nucl.\ Phys.\  {\bf 62}, 48 (2009)}
\arXiv{0807.3033}{[arXiv:0807.3033]};
%
U.~W.~Heinz,
\arXiv{0901.4355}{[arXiv:0901.4355]}.
%
\bibitem{strong_rdp}
R. D. Pisarski, 
Proc. Sci., LATTICE2008 (2008) 016
\arXiv{0810.4585}{[arXiv:0810.4585]}.
%
\bibitem{hydro}
P.~Romatschke,
\mydoi{10.1142/S0218301310014613}{\ijm{E 19}{1}{2010}}
\arXiv{0902.3663}{[arXiv:0902.3663]};
T.~Schafer and D.~Teaney,
\mydoi{10.1088/0034-4885/72/12/126001}{\ptps{72}{126001}{2009}}
\arXiv{0904.3107}{[arXiv:0904.3107]};
%
D.~A.~Teaney,
\arXiv{0905.2433}{[arXiv:0905.2433]};
and references therein these works.
%
\bibitem{Csernai:2006zz}
  L.~P.~Csernai, J.~I.~Kapusta and L.~D.~McLerran,
  \mydoi{10.1103/PhysRevLett.97.152303}{\prl{97}{152303}{2006}}
  \arXiv{nucl-th/0604032}{[arXiv:nucl-th/0604032]}.
%
\bibitem{PhysicalKinetics}
L.~D.~Landau and E.~M.~Lifshitz,
{\it Physical Kinetics} (Pergamon  Press, NewYork, 1981).
%
\bibitem{KuboFormula}
R.~Kubo and K.~Tomita,
\mydoi{10.1143/JPSJ.9.888}{J.\ Phys.\ Soc.\ Jpn. {\bf 9}, 888 (1954)};
%
H.~Nakano,
\mydoi{10.1143/PTP.15.77}{Prog.\ Theor.\ Phys.\ {\bf 15}, 77 (1956)};
%
R.~Kubo,
\mydoi{10.1143/JPSJ.12.570}{J.\ Phys.\ Soc.\ Jpn.\ {\bf 12}, 570 (1957)}.
%
\bibitem{lattice1} 
  F.~Karsch and H.~W.~Wyld,
\mydoi{10.1103/PhysRevD.35.2518}{\prd{35}{2518}{1987}};
  A.~Nakamura and S.~Sakai,
\mydoi{10.1103/PhysRevLett.94.072305}{\prl{94}{072305}{2005}}
\arXiv{hep-lat/0406009}{[arXiv:hep-lat/0406009]};
%
H.~B.~Meyer,
\mydoi{10.1103/PhysRevD.76.101701}{\prd{76}{101701}{2007}}
\arXiv{0704.1801}{[arXiv:0704.1801]};
\mydoi{10.1103/PhysRevLett.100.162001}{\prl{100}{162001}{2008}}
\arXiv{0710.3717}{[arXiv:0710.3717]};
%
  D.~Kharzeev and K.~Tuchin,
\mydoi{10.1088/1126-6708/2008/09/093}{\jhep{0809}{093}{2008}}
\arXiv{0705.4280}{[arXiv:0705.4280]};
  F.~Karsch, D.~Kharzeev and K.~Tuchin,
\mydoi{10.1016/j.physletb.2008.01.080}{\plb{663}{217}{2008}}
\arXiv{0711.0914}{[arXiv:0711.0914]}.
%
\bibitem{Huebner:2008as}
  K.~Huebner, F.~Karsch and C.~Pica,
\mydoi{10.1103/PhysRevD.78.094501}{\prd{78}{094501}{2008}}
\arXiv{0808.1127}{[arXiv:0808.1127]}.
%
\bibitem{Meyer:2009jp}
As a latest survey of the lattice calculations of tranport coefficients,
see 
  H.~B.~Meyer,
  \mydoi{10.1016/j.nuclphysa.2009.09.053}{\npa{830}{641C}{2009}}
  \arXiv{0907.4095}{[arXiv:0907.4095]}.
%
 %
\bibitem{susy1}
G.~Policastro, D.~T.~Son and A.~O.~Starinets,
\mydoi{10.1103/PhysRevLett.87.081601}{\prl{87}{081601}{2001}}
\arXiv{hep-th/0104066}{[arXiv:hep-th/0104066]};
\mydoi{10.1088/1126-6708/2002/09/043}{\jhep{0209}{043}{2002}}
\arXiv{hep-th/0205052}{[arXiv:hep-th/0205052]};
  A.~Buchel and J.~T.~Liu,
\mydoi{10.1103/PhysRevLett.93.090602}{\prl{93}{090602}{2004}}
  \arXiv{hep-th/0311175}{[arXiv:hep-th/0311175]};
%
P.~Kovtun, D.~T.~Son and A.~O.~Starinets,
\mydoi{10.1103/PhysRevLett.94.111601}{\ibid{94}{111601}{2005}}
\arXiv{hep-th/0405231}{[arXiv:hep-th/0405231]};
D.~T.~Son and A.~O.~Starinets,
\mydoi{10.1146/annurev.nucl.57.090506.123120}{\arnps{57}{95}{2007}}
\arXiv{0704.0240}{[arXiv:0704.0240]};
  R.~Baier, P.~Romatschke, D.~T.~Son, A.~O.~Starinets and M.~A.~Stephanov,
\mydoi{10.1088/1126-6708/2008/04/100}{\jhep{0804}{100}{2008}}
  \arXiv{0712.2451}{[arXiv:0712.2451]};
%
S.~S.~Gubser and A.~Karch,
\mydoi{10.1146/annurev.nucl.010909.083602}{\arnps{59}{145}{2009}}
\arXiv{0901.0935}{[arXiv:0901.0935]}.
%
\bibitem{susy2}
S.~S.~Gubser and A.~Nellore,
\mydoi{10.1103/PhysRevD.78.086007}{\prd{78}{086007}{2008}}
\arXiv{0804.0434}{[arXiv:0804.0434]};
%
U.~Gursoy, E.~Kiritsis, L.~Mazzanti and F.~Nitti,
\mydoi{10.1103/PhysRevLett.101.181601}{\prl{101}{181601}{2008}}
\arXiv{0804.0899}{[arXiv:0804.0899]};
\mydoi{10.1016/j.nuclphysb.2009.05.017}{\npb{820}{148}{2009}}
\arXiv{0903.2859}{[arXiv:0903.2859]};
%
N.~Evans and E.~Threlfall,
\mydoi{10.1103/PhysRevD.78.105020}{\prd{78}{105020}{2008}}
\arXiv{0805.0956}{[arXiv:0805.0956]};
%
J.~Noronha, M.~Gyulassy and G.~Torrieri,
\mydoi{10.1103/PhysRevLett.102.102301}{\prl{102}{102301}{2009}}
\arXiv{0807.1038}{[arXiv:0807.1038]};
\arXiv{0906.4099}{[arXiv:0906.4099]};
%
S.~S.~Gubser, S.~S.~Pufu, F.~D.~Rocha and A.~Yarom,
\arXiv{0902.4041}{[arXiv:0902.4041]};
%
U.~Gursoy, E.~Kiritsis, G.~Michalogiorgakis and F.~Nitti,
\mydoi{10.1088/1126-6708/2009/12/056}{\jhep{0912}{056}{2009}}
\arXiv{0906.1890}{[arXiv:0906.1890]}.
%
 \bibitem{Jeon}
  S.~Jeon,
 \mydoi{10.1103/PhysRevD.52.3591}{\prd{52}{3591}{1995}}
\arXiv{hep-ph/9409250}{[arXiv:hep-ph/9409250]};
  S.~Jeon and L.~G.~Yaffe,
\mydoi{10.1103/PhysRevD.53.5799}{\ibid{53}{5799}{1996}}
\arXiv{hep-ph/9512263}{[arXiv:hep-ph/9512263]}.
%
\bibitem{diagram}
A.~Hosoya, M.~A.~Sakagami and M.~Takao,
\mydoi{10.1016/0003-4916(84)90144-1}{Annals Phys.\  {\bf 154}, 229 (1984)};
  E.~Wang and U.~W.~Heinz,
\mydoi{10.1016/S0370-2693(99)01324-6}{\plb{471}{208}{1999}}
  \arXiv{hep-ph/9910367}{[arXiv:hep-ph/9910367]};
  M.~E.~Carrington, D.~f.~Hou and R.~Kobes,
\mydoi{10.1103/PhysRevD.62.025010}{\prd{62}{025010}{2000}}
 \arXiv{hep-ph/9910344}{[arXiv:hep-ph/9910344]};
 M.~A.~Valle Basagoiti,
\mydoi{10.1103/PhysRevD.66.045005}{\ibid{66}{045005}{2002}}
  \arXiv{hep-ph/0204334}{[arXiv:hep-ph/0204334]};
  \mydoi{10.1103/PhysRevD.67.025022 }{\ibid{67}{025022}{2003}}
  \arXiv{hep-th/0201116}{[arXiv:hep-th/0201116]};
  H.~Defu,
\arXiv{hep-ph/0501284}{[arXiv:hep-ph/0501284]};
  D.~Fernandez-Fraile,
\arXiv{1009.2741}{[arXiv:1009.2741]}.

 \bibitem{2PI}
  G.~Aarts and J.~M.~Martinez Resco,
\mydoi{10.1103/PhysRevD.68.085009}{\prd{68}{085009}{2003}}
 \arXiv{hep-ph/0303216}{[arXiv:hep-ph/0303216]};
\mydoi{10.1088/1126-6708/2004/02/061}{\jhep{0402}{061}{2004}}
  \arXiv{hep-ph/0402192}{[arXiv:hep-ph/0402192]};
\mydoi{10.1088/1126-6708/2005/03/074}{\ibid{0503}{074}{2005}}
  \arXiv{hep-ph/0503161}{[arXiv:hep-ph/0503161]};
%
  M.~E.~Carrington and E.~Kovalchuk,
\mydoi{10.1103/PhysRevD.76.045019 ]}{\prd{76}{045019}{2007}}
 \arXiv{0705.0162}{[arXiv:0705.0162]}.

\bibitem{3PI}
  M.~E.~Carrington and E.~Kovalchuk,
\mydoi{10.1103/PhysRevD.77.025015}{\prd{77}{025015}{2008}}
  \arXiv{0709.0706}{[arXiv:0709.0706]};
\mydoi{10.1103/PhysRevD.80.085013}{\ibid{80}{085013}{2009}}
 \arXiv{0906.1140}{[arXiv:0906.1140]}.

\bibitem{Gagnon}
  J.~S.~Gagnon and S.~Jeon,
\mydoi{10.1103/PhysRevD.75.025014}{\prd{75}{025014}{2007}}
 \mydoi{10.1103/PhysRevD.76.089902}{[Erratum-\ibid{76}{089902}{2007}]}
\arXiv{hep-ph/0610235}{[arXiv:hep-ph/0610235]};
  \mydoi{10.1103/PhysRevD.76.105019}{\ibid{76}{105019}{2007}}
\arXiv{0708.1631}{[arXiv:0708.1631]}.

 
\bibitem{hadronic}
  S.~Gavin,
\mydoi{10.1016/0375-9474(85)90190-3}{\npa{435}{826}{1985}};
%
  M.~Prakash, M.~Prakash, R.~Venugopalan and G.~Welke,
\mydoi{10.1016/0370-1573(93)90092-R}{\phr{227}{321}{1993}};
%
  D.~Davesne,
\mydoi{10.1103/PhysRevC.53.3069}{\prc{53}{3069}{1996}};
%
  A.~Dobado and S.~N.~Santalla,
\mydoi{10.1103/PhysRevD.65.096011}{\prd{65}{096011}{2002}}
  \arXiv{hep-ph/0112299}{[arXiv:hep-ph/0112299]};
 %
  A.~Dobado and F.~J.~Llanes-Estrada,
\mydoi{10.1103/PhysRevD.69.116004}{\ibid{69}{116004}{2004}}
\arXiv{hep-ph/0309324}{[arXiv:hep-ph/0309324]};
%
  D.~Fernandez-Fraile and A.~Gomez Nicola,
\mydoi{10.1103/PhysRevD.73.045025}{\ibid{73}{045025}{2006}}
  \arXiv{hep-ph/0512283}{[arXiv:hep-ph/0512283]};
  J.~W.~Chen and E.~Nakano,
\mydoi{10.1016/j.physletb.2007.02.026}{\plb{647}{371}{2007}}
  \arXiv{hep-ph/0604138}{[arXiv:hep-ph/0604138]};
  J.~W.~Chen, Y.~H.~Li, Y.~F.~Liu and E.~Nakano,
\mydoi{10.1103/PhysRevD.76.114011}{\prd{76}{114011}{2007}}
\arXiv{hep-ph/0703230}{[arXiv:hep-ph/0703230]};
  K.~Itakura, O.~Morimatsu and H.~Otomo,
\mydoi{10.1103/PhysRevD.77.014014}{\ibid{ 77}{014014}{2008}}
\arXiv{0711.1034}{[arXiv:0711.1034]};
  J.~W.~Chen and J.~Wang,
\mydoi{10.1103/PhysRevC.79.044913}{\prc{79}{044913}{2009}}
\arXiv{0711.4824}{[arXiv:0711.4824]};
  D.~Fernandez-Fraile and A.~G.~Nicola,
\mydoi{10.1103/PhysRevLett.102.121601}{\prl{102}{121601}{2009}}
  \arXiv{0809.4663}{[arXiv:0809.4663]};
  C.~Sasaki and K.~Redlich,
\mydoi{10.1103/PhysRevC.79.055207}{\prc{79}{055207}{2009}}
\arXiv{0806.4745}{[arXiv:0806.4745]};
  \mydoi{10.1016/j.nuclphysa.2009.11.005}{\npa{832}{62}{2010}}
\arXiv{0811.4708}{[arXiv:0811.4708]};
  P.~Chakraborty and J.~I.~Kapusta,
\arXiv{1006.0257}{[arXiv:1006.0257]}.
\bibitem{transport}
U.~W.~Heinz,
\mydoi{10.1103/PhysRevLett.51.351}{\prl{51}{351}{1983}};
\mydoi{10.1016/0003-4916(85)90336-7}{\anp{161}{48}{1985}};
\mydoi{10.1016/0003-4916(86)90114-4}{\ibid{168}{148}{1986}}.
%
A.~Hosoya and K.~Kajantie,
\mydoi{10.1016/0550-3213(85)90499-7}{\npb{250}{666}{1985}}; 
G.~Baym, H.~Monien, C.~J.~Pethick and D.~G.~Ravenhall,
\mydoi{10.1103/PhysRevLett.64.1867}{\prl{64}{1867}{1990}};
%
M.~H.~Thoma,
\mydoi{10.1016/0370-2693(91)91466-9}{\plb{269}{144}{1991}};
D.~W.~von Oertzen,
\mydoi{10.1016/0370-2693(92)90780-8}{\ibid{280}{103}{1992}};
H.~Heiselberg,
\mydoi{10.1103/PhysRevD.49.4739}{\prd{49}{4739}{1994}}
\arXiv{hep-ph/9401309}{[arXiv:hep-ph/9401309]};
\mydoi{10.1103/PhysRevLett.72.3013}{\prl{72}{3013}{1994}}
\arXiv{hep-ph/9401317}{[arXiv:hep-ph/9401317]};
G.~Baym and H.~Heiselberg,
\mydoi{10.1103/PhysRevD.56.5254}{\prd{56}{5254}{1997}}
\arXiv{astro-ph/9704214}{[arXiv:astro-ph/9704214]};
J.~Ahonen,
\mydoi{10.1103/PhysRevD.59.023004}{\ibid{59}{023004}{1999}}
\arXiv{hep-ph/9801434}{[arXiv:hep-ph/9801434]};
%
  M.~Asakawa, S.~A.~Bass and B.~Muller,
\mydoi{10.1103/PhysRevLett.96.252301}{\prl{96}{252301}{2006}}
  \arXiv{hep-ph/0603092}{[arXiv:hep-ph/0603092]};
\mydoi{10.1143/PTP.116.725}{\ptps{116}{725}{2007}}
  \arXiv{hep-ph/0608270}{[arXiv:hep-ph/0608270]};
%
  Z.~Xu and C.~Greiner,
\mydoi{10.1103/PhysRevLett.100.172301}{\prl{100}{172301}{2008}}
  \arXiv{0710.5719}{[arXiv:0710.5719]};
  M.~A.~York and G.~D.~Moore,
\mydoi{10.1103/PhysRevD.79.054011}{\prd{79}{054011}{2009}}
 \arXiv{0811.0729}{[arXiv:0811.0729]};
  J.~W.~Chen, H.~Dong, K.~Ohnishi and Q.~Wang,
  \mydoi{10.1016/j.physletb.2010.01.072}{\plb{685}{277}{2010}}
\arXiv{0907.2486}{[arXiv:0907.2486]}.
 %
 \bibitem{amy}
P.~Arnold, G.~D.~Moore and L.~G.~Yaffe,
\mydoi{10.1088/1126-6708/2000/11/001}{\jhep{0011}{001}{2000}}
\arXiv{hep-ph/0010177}{[arXiv:hep-ph/0010177]};
\mydoi{10.1088/1126-6708/2003/05/051}{{\it ibid} {\bf 0305}, 051 (2003)}
\arXiv{hep-ph/0302165}{[arXiv:hep-ph/0302165]}.
%
\bibitem{semiQGP}
  Y.~Hidaka and R.~D.~Pisarski,
\mydoi{10.1103/PhysRevD.78.071501}{\prd{78}{071501}{2008}}
\arXiv{0803.0453}{[arXiv:0803.0453]};
 \mydoi{10.1103/PhysRevD.80.036004}{\ibid{80}{036004}{2009}}
\arXiv{0907.4609}{[arXiv:0907.4609]};
 \mydoi{10.1103/PhysRevD.80.074504}{\ibid{80}{074504}{2009}}
\arXiv{0907.4609}{[arXiv:0907.4609]};
\mydoi{10.1103/PhysRevD.81.076002}{\ibid{81}{076002}{2010}}
\arXiv{0912.0940}{[arXiv:0912.0940]}.
%
\bibitem{nextLeading1}
  G.~D.~Moore,
\mydoi{10.1103/PhysRevD.76.107702}{\prd{76}{107702}{2007}}
\arXiv{0706.3692}{[arXiv:0706.3692]}.
\bibitem{nextLeading2}
  M.~E.~Carrington and E.~Kovalchuk,
  \mydoi{10.1103/PhysRevD.81.065017}{\ibid{81}{065017}{2010}}
  \arXiv{0912.3149}{[arXiv:0912.3149]}.
%
\bibitem{Eliashberg:1962}
G.~M.~Eliashberg, 
Sov. Phys. JETP {\bf 14}, 886 (1962).
%
 \bibitem{lebellac}
M. Le Bellac, {\it Thermal Field Theory} (Cambridge University Press,
Cambridge, 2000).
%
\bibitem{Aurenche:1991hi}
  P.~Aurenche and T.~Becherrawy,
\mydoi{10.1016/0550-3213(92)90597-5}{\npb{379}{259}{1992}}.
%
\bibitem{vanEijck:1994rw}
  M.~A.~van Eijck, R.~Kobes and C.~G.~van Weert,
 \mydoi{10.1103/PhysRevD.50.4097}{\prd{50}{4097}{1994}}
  \arXiv{hep-ph/9406214}{[arXiv:hep-ph/9406214]}.
%
\bibitem{Gelis:2001xt}
  F.~Gelis, D.~Schiff and J.~Serreau,
  \mydoi{10.1103/PhysRevD.64.056006 }{\prd{64}{056006}{2001}}
  \arXiv{hep-ph/0104075}{[arXiv:hep-ph/0104075]}.
%
\bibitem{linearizedBoltzmann}
S.~Chapman and T.~G.~Cowling, 
 {\it The Mathematical Theory of Non-Uniform Gases (3rd ed.)}  (Cambridge, 1970).
%
\bibitem{Koide:2009sy}
  T.~Koide, E.~Nakano and T.~Kodama,
\mydoi{10.1103/PhysRevLett.103.052301}{\prl{103}{052301}{2009}}
  \arXiv{0901.3707}{[arXiv:0901.3707]}.
%
\bibitem{Altherr:1989rk}
  T.~Altherr,
  \mydoi{10.1016/0370-2693(90)91749-2}{Phys.\ Lett.\  B {\bf 238}, 360 (1990)}.
%
\bibitem{hidaka:2011}
 Y.~Hidaka and T.~Kunihiro, in preparation.  
%
 \bibitem{LPM}
L.~D.~Landau and I.~Pomeranchuk, Dokl. Akad. Nauk Ser. Fiz. {\bf 92}  
535 (1953); \ibid{92}{735}{1953};
A.~B.~Migdal, Dokl. Akad. Nauk S.S.S.R. {\bf 105}, 77 (1955); 
\mydoi{10.1103/PhysRev.103.1811}{\pr{103}{1811}{1956}}.
%
 \bibitem{Arnold:2002zm}
  P.~B.~Arnold, G.~D.~Moore and L.~G.~Yaffe,
\mydoi{10.1088/1126-6708/2003/01/030}{JHEP {\bf 0301}, 030 (2003)}
\arXiv{hep-ph/0209353}{[arXiv:hep-ph/0209353]}.
%
\bibitem{Kobes:1985kc}
  R.~L.~Kobes and G.~W.~Semenoff,
 \mydoi{10.1016/0550-3213(86)90006-4}{\npb{260}{714}{1985}};
  \mydoi{10.1016/0550-3213(85)90056-2}{\ibid{272}{329}{1986}}.
%
\bibitem{Toyoda:1986jn}
  T.~Toyoda,
\mydoi{10.1016/0003-4916(87)90100-X}{\anp{173}{226}{1986}}.
  %
\bibitem{Aarts:2002tn}
  G.~Aarts and J.~M.~Martinez Resco,
\mydoi{10.1088/1126-6708/2002/11/022}{\jhep{0211}{022}{2002}}
 \arXiv{hep-ph/0209048}{[arXiv:hep-ph/0209048]}.
\end{thebibliography}
\end{document}